%% file: ms.tex
\documentclass[apj]{emulateapj}  
\pdfoutput=1
\usepackage{apjfonts}
\usepackage{amsmath}
\usepackage[usenames]{color}

\usepackage[colorlinks=true,
  citecolor=blue,  
  frenchlinks=True,
  breaklinks]{hyperref}
\usepackage{rotating}  

\usepackage{float}
\usepackage{epsfig}
\usepackage{comment}
\usepackage{mathtools}

\shorttitle{The properties of \Ha\ ELGs at $z$ = 2.24}
\shortauthors{An et al.}

\begin{document}
\definecolor{purple}{RGB}{160,32,240}
\newcommand{\peter}[1]{}
\newcommand{\fa}{F_\mathrm{H\alpha}}
\newcommand{\hinv}{h^{-1}}
\newcommand{\mpc}{\rm{Mpc}}
\newcommand{\hmpc}{$\hinv\mpc$}

\newcommand{\sfr}{\mathrm{SFR}}
\newcommand{\Msun}{M_{\odot}}
\newcommand{\Ha}{H$\alpha$}
\newcommand{\OIII}{[{O}~{\scriptsize {III}}]}
\newcommand{\NII}{[{N}~{\scriptsize {II}}]}
\newcommand{\FeII}{[{Fe}~{\scriptsize {II}}]}
\newcommand{\OII}{[{O}~{\scriptsize {II}}]}
\newcommand{\SIII}{[{S}~{\scriptsize {III}}]}
\newcommand{\SII}{[{S}~{\scriptsize {II}}]}
\newcommand{\C}{$\kappa({\rm L})$}
\newcommand{\mm}{$\mu$m}
\newcommand\aper[1]{#1\arcsec\ diameter}%
\newcommand{\Hb}{H$\beta$}

\newcommand{\phistar}{\Phi^{\star}}
\newcommand{\Lstar}{L^{\star}}
\newcommand{\Mstar}{M^{\star}}
\newcommand{\SFRd}{\rho_{\rm SFR}}
\newcommand{\iyr}{yr$^{-1}$}
\newcommand{\vMpc}{Mpc$^{-3}$}

\newcommand{\fluxunit}{erg s$^{-1}$ cm$^{-2}$}

\newcommand{\mumod}{\langle$log\,(EW$_{\rm 0}/$\AA)$\rangle}

\title{THE PROPERTIES OF \Ha\ EMISSION-LINE GALAXIES AT $z$ = 2.24$^{\dagger}$}

\author{Fang~Xia An\altaffilmark{1}, Xian~Zhong Zheng\altaffilmark{1}, Wei-Hao Wang\altaffilmark{2}, Jia-Sheng Huang\altaffilmark{3,4}, Xu Kong\altaffilmark{5}, Jun-Xian Wang\altaffilmark{5}, Guan~Wen Fang\altaffilmark{6,7}, Feifan Zhu\altaffilmark{5},
Qiu-Sheng Gu\altaffilmark{8}, Hong Wu\altaffilmark{4}, Lei Hao\altaffilmark{9}, And Xiao-Yang Xia\altaffilmark{10}}

\altaffiltext{1}{Purple Mountain Observatory, China Academy of Sciences, West Beijing Road 2, Nanjing 210008, China; fangxiaan@pmo.ac.cn}
\altaffiltext{2}{Academia Sinica Institute of Astronomy and Astrophysics,Taipei 10617, Taiwan}
\altaffiltext{3}{Harvard-Smithsonian Center for Astrophysics, 60 Garden Street, Cambridge, MA 02138, USA}
\altaffiltext{4}{National Astronomical Observatories, Chinese Academy of Sciences, Beijing 100012, China}
\altaffiltext{5}{CAS Key Laboratory for Research in Galaxies and Cosmology, Department of Astronomy, University of Science and Technology of China, Hefei 230026, China}
\altaffiltext{6}{Institute for Astronomy and History of Science and Technology, Dali University, Yunnan 671003, China}
\altaffiltext{7}{Key Laboratory of Modern Astronomy and Astrophysics, Nanjing University, Ministry of Education, Nanjing 210093, China}
\altaffiltext{8}{School of Astronomy and Space Science, Nanjing University, 22 Hankou RD, Nanjing 210093, China}
\altaffiltext{9}{Shanghai Astronomical Observatory, Chinese Academy of Sciences, 80 Nandan Road, Shanghai 2000
30, China}
\altaffiltext{10}{Tianjin Astrophysics Center, Tianjin Normal University, Tianjin 300387, China}

\altaffiltext{$\dagger$}{Based on observations obtained with WIRCam, a joint project of CFHT, Taiwan, Korea, Canada, France, and the Canada-France-Hawaii Telescope (CFHT) which is operated by the National Research Council (NRC) of Canada, the Institute National des Sciences de l'Univers of the Centre National de la Recherche Scientifique of France, and the University of Hawaii.}

\begin{abstract}
Using deep narrow-band $H_2S1$ and $K_{\rm s}$-band imaging data obtained with CFHT/WIRCam, we identify a sample of 56 \Ha\ emission-line galaxies (ELGs) at $z=2.24$ with the 5$\sigma$ depths of $H_2S1=22.8$ and $K_{\rm s}=24.8$ (AB) over 383 arcmin$^{2}$ area in the Extended Chandra Deep Field South. 
A detailed analysis is carried out with existing multi-wavelength data in this field. 
Three of the 56 \Ha\ ELGs are detected in \textit{Chandra} 4\,Ms X-ray observations and two of them are classified as active galactic nuclei.
The rest-frame UV and optical morphologies revealed by \textit{HST}/ACS and WFC3 deep images show that nearly half of the \Ha\ ELGs are either merging systems or have a close companion, indicating that the merging/interacting processes play a key role in regulating star formation at cosmic epoch $z=2-3$;
About 14\% are too faint to be resolved in the rest-frame UV morphology due to high dust extinction.
We estimate dust extinction from spectral energy distributions.
We find that dust extinction is generally correlated with \Ha\ luminosity and stellar mass.
Our results suggest that \Ha\ ELGs are representative of star-forming galaxies.
Applying extinction corrections for individual objects, we examine the intrinsic \Ha\ luminosity function (LF) at $z=2.24$, obtaining a best-fit Schechter function characterized by a faint-end slope of $\alpha=-1.3$.
This is shallower than the typical slope of $\alpha=\sim -1.6$ in previous works based on constant extinction correction.
We demonstrate that this difference is mainly due to the different extinction corrections.
The proper extinction correction is thus the key to recovering the intrinsic LF as the extinction globally increases with \Ha\ luminosity. 
Moreover, we find that our \Ha\ LF mirrors the stellar mass function of star-forming galaxies at the same cosmic epoch.
This finding indeed reflects the tight correlation between star formation rate and stellar mass for the star-forming galaxies, i.e., the so-called main sequence.

\end{abstract}

\section{INTRODUCTION}\label{s:introduction}

The cosmic star formation rate (SFR) density peaks at $z\sim 2-3$ and rapidly declines to the present day \citep[][and references therein]{Hopkins06,Karim11,Sobral13}.
Systematic studies of galaxy populations at the peak epoch hold the key to our understanding of galaxy formation and evolution, in particular for massive galaxies.
The population of massive quiescent galaxies began to emerge \citep[e.g.,][]{Brammer11} and the bulk of stars in local massive galaxies were formed at $z>\sim 1.5$ \citep{Renzini06}. 
A number of techniques have been developed to probe different galaxy populations in this crucial epoch, including Lyman break selection \citep{Steidel99}, red color cut \citep{Franx03}, $BzK$ selection \citep{Daddi04}, submillimeter detection \citep{Chapman05} and narrow-band imaging (\citealt{Moorwood00}; see \citealt{Shapley11} for a review).
A complete picture is still missing due to the short of spectroscopic follow ups in the ``redshift desert'' $1.4<z<2.5$ for large samples selected in terms of physical quantities such as stellar mass or SFR. 

The deep near-infrared (NIR) observations play a central role in probing high-$z$ galaxies in the rest-frame optical, which is more reflective of older stars (hence stellar mass). 
The NIR narrow-band imaging has turned out to be a modest way to identify emission-line galaxies (ELGs) in a narrow redshift range ($\delta z/(1+z)=1\%-2$\%) over large sky coverage \citep{Sobral13}.
It provides redshifts with higher precision than the photometric redshift method based on broad-band spectral energy distributions (SEDs), and allows us to distinguish environments (e.g., groups or clusters) traced by the ELGs \citep[e.g.,][]{Matsuda11,Hatch11,Geach12}. 
Moreover, the strength of an emission line, such as \Ha\ and \OII $\lambda$3727, can be measured by the flux excess in the narrow-band with respect to the corresponding broad-band.
The optical and NIR emission lines from ionized gas surrounding massive young stars represent a nearly instantaneous measure of the SFR \citep{Kennicutt12}.
The \Ha\ is often used as an SFR indicator because it is  the strongest emission line in the optical/NIR, and less affected by dust obscuration compared to emission lines in the UV such as Ly$\alpha$ in typical star-forming galaxies (SFGs). 

The past decade has seen a number of NIR narrow-band surveys using ground-based telescopes \citep[e.g.,][]{Geach08,Ly11} and NIR grism spectroscopic surveys with the \textit{Hubble Space Telescope} \citep[\textit{HST}; e.g.,][]{Yan99,Atek10,vanDokkum11,Colbert13} in the study of \Ha\ ELGs at $z>$0.4, yielding \Ha\ luminosity functions (LFs) with a steep faint-end slope $\alpha\sim -1.6$ out to $z\sim 2.5$ \citep{Hayes10,Ly11,Sobral13}.
The derived \Ha\ LFs suffer from large uncertainties mostly arising from poor extinction correction due to the lack of ancillary data.
Detailed studies of the properties of \Ha\ ELGs will help to reduce the uncertainties.
On the other hand, SFGs are found to follow a tight correlation between SFR and stellar mass \citep[e.g.,][]{Noeske07,Daddi07}.
This correlation, namely the main sequence, has been convincingly established from the local universe \citep{Brinchmann04} to high-redshift universe \citep{Guo13,Peng10,Karim11,Reddy12}.
One would expect that the SFR function traces the stellar mass function (SMF) for SFGs. 
Since \Ha\ luminosity is a direct measure of SFR, the \Ha\ LF is expected to mirror the SMF.  
However, the SMF tends to have a shallow faint-end with a slope $\alpha\sim -1.3$ over a wide redshift range $0<z<3$ \citep[e.g.,][]{Ilbert10,Brammer11}.
The cause of this discrepancy in the faint-end slope between the two functions remains to be explored. 

We carry out a deep 2.13\,$\micron$ narrow-band imaging survey to search for \Ha\ ELGs at $z$ = 2.24 in the Extended Chandra Deep Field South (ECDFS).
This field contains the deepest observations from \textit{Chandra}, \textit{Galaxy Evolution Explorer}, \textit{HST}, and \textit{Spitzer}.
With these data, we are able to securely identify a sample of \Ha\ ELGs, perform a detailed analysis and determine the intrinsic \Ha\ LF.
We show that resolving dusty SFGs with intrinsically luminous \Ha\ from those with observed faint \Ha\ leads to a shallow faint-end slope for the \Ha\ LF.
We describe observations and data reduction in Section~\ref{s:observation}.
Section~\ref{s:Sample} gives the selection of \Ha\ ELGs and SED analysis to obtain photometric redshift, stellar mass and extinction.
We present the properties of \Ha\ ELGs in Section~\ref{s:property}.
The derived \Ha\ LF at $z=2.24$ is given in Section~\ref{s:Results}.
We discuss and summarize our results in Section~\ref{s:discussion}.
Throughout the paper we adopt a cosmology with [$\Omega_{\Lambda}$, $\Omega_M$, $h_{70}$] = [0.7, 0.3, 1.0].
Kroupa initial mass function \citep[IMF; ][]{Kroupa01} and the AB magnitude system \citep{Oke74} are used unless otherwise stated.

\section{OBSERVATIONS AND DATA REDUCTION} \label{s:observation}
The deep narrow-band imaging of the ECDFS ($\alpha$ = 03:28:45, $\delta$ = $-$27:48:00) were taken with WIRCam on board the Canada-France-Hawaii Telescope \citep[CFHT; ][]{Puget04} through the $H_2S1$ filter ($\lambda_\mathrm{c}=2.130\,\micron$, $\Delta \lambda=0.0293$\,$\micron$).
WIRCam is equipped with four 2048$\times$2048 HAWAII2-RG detectors, providing a field of view 20$\arcmin\times 20\arcmin$ and a pixel scale of 0$\farcs$3\, pixel$^{-1}$.
The gaps between detectors are 45$\arcsec$.
The $H_2S1$ observations were carried out under seeing conditions of 0$\farcs$6$-0\farcs$8 with a total integration time of 17.22\,hrs in semester 2011B.
Dithering technique was adopted in observations in order to remove bad pixels and cover gaps between detectors.

The data were reduced using the pipeline \textit{SIMPLE} (Simple Imaging and Mosaicking Pipeline) written in IDL \citep{Wang10,Hsieh12}, which include flat-fielding, background subtraction, removing cosmic rays, and instrumental features like crosstalk and residual images from saturated objects in previous exposures and masking satellite trails. 
Because of the rapid variation of sky background in the NIR, only exposures taken in the same dithering block (within 40 minutes) and by the same detector are reduced in the same run and combined into a background-subtracted science image. 
After that, the science images of four detectors are mosaicked into a frame science image.
The $H_2S1$ images are astrometrically calibrated with the \textit{HST} observations from the Galaxy Evolution from Morphology and SEDs \citep[GEMS;][]{Rix04,Caldwell08} survey using a list of 7415 compact sources with $K_\mathrm{s}<21.5$.
The accuracy of astrometry is $\sim 0\farcs$1.
The final science image was produced by co-adding 326 frame science images.
The exposure time map was also created in the co-adding.
The photometric calibration is performed with the $K_\mathrm{s}$-band photometric catalog of the Great Observatories Origins Deep Survey (GOODS) South extracted from VLT/ISAAC observations \citep{Retzlaff10}.
In total 125 point sources with $16.0<K_\mathrm{s}<19.0$ are chosen as photometric standard stars used to build empirical point spread function (PSF).
A correction of 1.26 is derived from the empirical PSF to convert a flux integrated within an aperture of diameter of 2$\arcsec$ into the total flux.
The calibrated fluxes of the chosen point sources in our mosaic $H_2S1$ image only show 1\% scatter compared with these given in the reference catalog. 
The final $H_2S1$ science image has 383 arcmin$^{2}$ area with a total integrated exposure time $>10$\,hr (and $>15$\,hr for 47.3\%).
We limited source detection in this area, where a 5$\sigma$ depth corresponds to 22.8 mag for point sources.

We also utilize $K_\mathrm{s}$-band ($\lambda_\mathrm{c}=2.146\,\micron$, $\Delta\lambda=0.325\,\micron$) imaging of ECDFS obtained with CFHT/WIRCam \citep{Hsieh12}.
The $K_\mathrm{s}$-band observations were taken with CFHT/WIRCam in semesters 2009B and 2010B.
The data were reduced and calibrated in the same way as we describe above for the $H_2S1$ data.
The $K_\mathrm{s}$ image reaches a 5$\sigma$ depth of $K_\mathrm{s}$ = 24.8\,mag for point sources in the $H_2S1$ source detection region. 
More technical details can be found in \citet{Wang10} and \citet{Hsieh12}.
The 5$\sigma$ limit for extended sources with a S{\'{e}}rsic index of unity is 0.4 mag lower for the $H_2S1$ image  and 0.3 mag lower for the $K_\mathrm{s}$ image.

\section{SAMPLE SELECTION} \label{s:Sample}

\subsection{Selection of Emission-line Objects}

We use SExtractor software \citep{Bertin96} to detect sources and measure fluxes in the $H_2S1$ image.
A minimum of five contiguous pixels with fluxes $> 2.5\sigma$ of the background noise is required for a secure source detection.
The exposure map is taken as weight image to reduce spurious detections in low signal-to-noise (S/N) regions.
The ``dual-image'' mode in SExtractor is used to perform photometry in the $K_\mathrm{s}$ image, namely, to measure fluxes in the $K_\mathrm{s}$ image over the same area of a detection as in the $H_2S1$ image.
Of course, the two images are well aligned into the same frame.
In total , 8720 sources are securely detected with an S/N $>5$ in both images.

\begin{figure}
\centering
\includegraphics[width=0.45\textwidth]{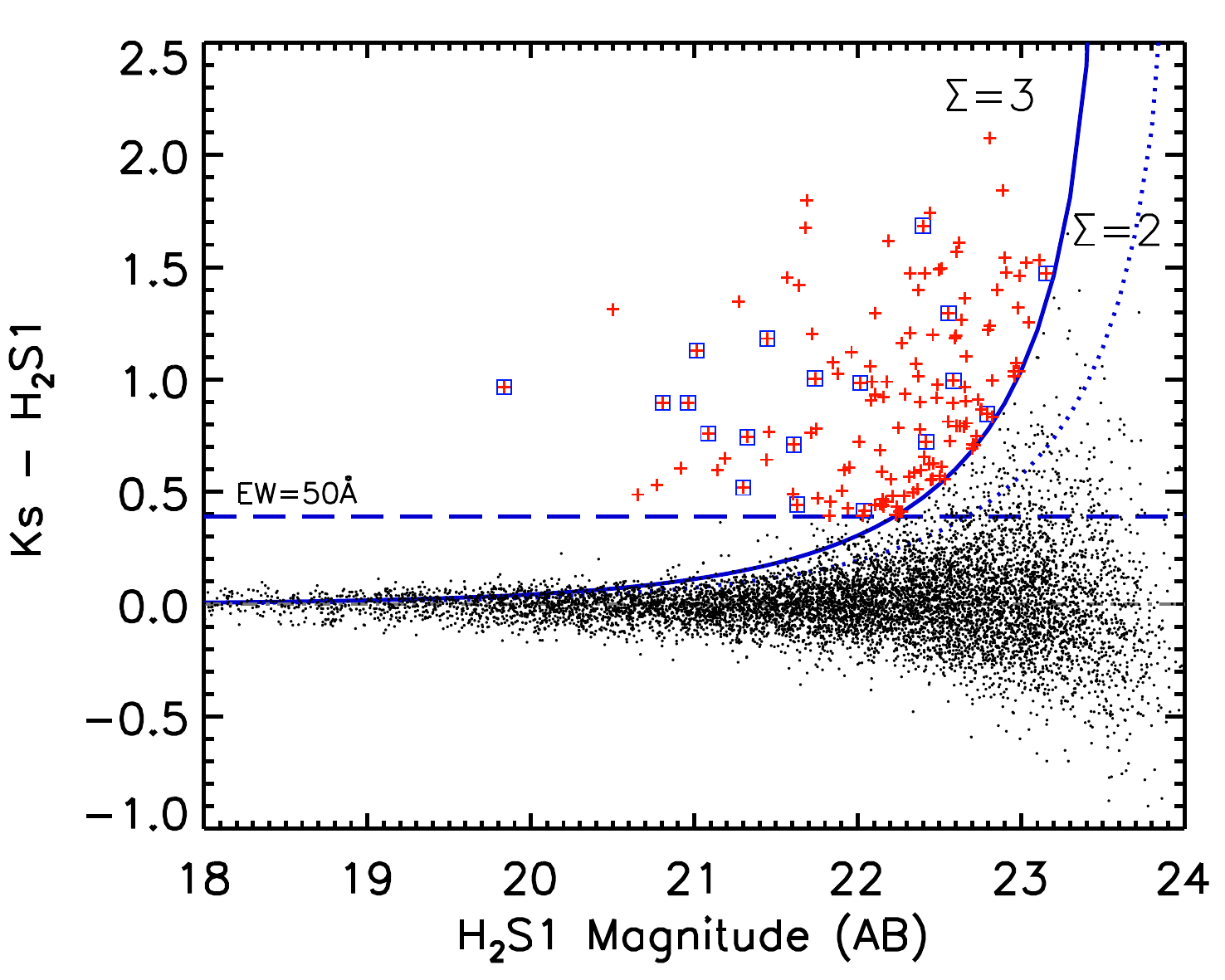}
\caption{Color-magnitude diagram for $H_\mathrm{2}S1$ detected objects.
The blue solid/dotted line represents the limitation due to background noise at the $\Sigma=3$/$\Sigma=2$ level.
The blue dashed line marks the excess cut corresponding to the rest-frame EW=50$\AA$.
The red crosses are the 140 emission-line candidates.
The blue squares mark the candidates with spec-$z$.}
\label{f:cdfs_mag_sele}
\end{figure}

As we have stated in Section~\ref{s:introduction}, the narrow-band excess is an emission-line indicator that is widely used to identify emission-line objects.
Here we use $K_{\rm s}-H_2S1$ to probe \Ha\ emission-line objects following \cite{Bunker95}.
A secure and significant narrow-band excess is mainly determined by the background noises of the narrow- and broad-band images $\sigma_{\rm H_\mathrm{2}S1}$ and $\sigma_{K_{\rm s}}$ as follows: 
\begin{equation}
K_{\rm s}-H_\mathrm{2}S1 > \Sigma \sqrt{\sigma_{K_{\rm s}}^2+\sigma_{\rm H_\mathrm{2}S1}^2},
\end{equation}
where the right-hand side term is the combined background noise of the two bands and $\Sigma$ is the significant factor. 
Figure~\ref{f:cdfs_mag_sele} shows the $K_{\rm s} - H_2S1$ color as a function of $H_2S1$ magnitude for 8720 sources.
The solid curve represents the combined noise with $\Sigma$ = 3 and the dotted curve refers to the noise with $\Sigma$ = 2.
We adopt $\Sigma>3$ to select emission-line candidates.
In order to minimize false excess caused by photon noises of bright objects, we also employ an empirical cut EW $>50$\,\AA\ following \cite{Geach08}, where EW refers to the rest-frame equivalent width of an emission line.
This cut corresponds to $K_{\rm s}-H_\mathrm{2}S1>0.39$\,mag.
We note that a lower EW cut (e.g., EW>30\,\AA) only increases a few more candidates and will not significantly change on our results.

As shown in Figure~\ref{f:cdfs_mag_sele}, 146 emission-line candidates are selected with $\Sigma>3$ and EW $>50$\,\AA.
To examine how many spurious sources will be selected by our criteria, we perform a ``negative'' selection with $K_{\rm s}-H_\mathrm{2}S1<-0.39$\,mag and $-3\Sigma$ cut.
Only one object being selected means our criteria enable a clean selection of emission-line object candidates.
Furthermore, we perform a visual examination of our selected candidates.
Of the 146, six are visually identified as spurious sources like spikes of bright stars or contaminations that were not rejected by data reduction.
The other 140 candidates are expected to be objects with a strong emission line between Pa$\alpha$ at $z=0.14$ and Ly$\alpha$ at $z=16.52$.
And 33 of the 140 objects are found to have spectroscopic redshifts from the catalog collected by \cite{Cardamone10}.

It is worth noting that the requirement of a secure detection ($> 5\sigma$) in the $K_{\rm s}$ might miss emission-line objects with very high EWs. 
To quantify this effect, we re-examine  $H_2S1$ sources and find 14 objects with $K_{\rm s}\geq $ 24.8\,mag. 
Among them, six are visually identified as spurious sources.
The other eight sources are very faint ($H_2S1 >$22.5\,mag), compared to the detection limit of $H_2S1 =$ 22.8\,mag. 
We suspect that their very high EWs ($> 500$\AA ) would be dominated by noise and the intrinsic EWs could be lower.
Considering that \Ha\ emitters represent  $30\%-40\%$ of all emitters \citep{Hayes10, Lee12}, less than three of the eight sources are expected to be \Ha\ emitters. 
Moreover, they must be very low-mass galaxies with given $K_{\rm s}$ magnitudes. 
The missing fraction of  \Ha\ emitters with very high EWs is estimated to be very small ($<$ 5\%). 
We therefore conclude that inclusion of the selection criterion $K_{\rm s}<$ 24.8\,mag  has negligible detrimental effects on our main results.

\subsection{Photometric Redshifts and SED Modeling} \label{s:Photometric Redshifts}

To measure fluxes and construct SEDs for the selected 140 emission-line objects, we use 12 bands imaging data, including $U$, $B$, $V$, $R$ and $I$-band data from the Multiwavelength Survey by Yale-Chile \citep[MUSYC; ][]{Gawiser06}, \textit{HST}/ACS F606W ($V_{606}$) and F850LP ($z_{850}$) imaging from GEMS \citep[]{Rix04,Caldwell08}, \textit{HST}/WFC3 F125W ($J_{125}$) and F160W ($H_{160}$) imaging from the Cosmic Assembly Near-infrared Deep Extragalactic Legacy Survey \citep[CANDELS;][]{Grogin11,Koekemoer11}, and CFHT/WIRCam $J$ and $K_{\rm s}$ imaging \citep{Hsieh12} in combination with our $H_2S1$ data.
Only 72 of the 140 have $J_{125}$ and $H_{160}$ data because CANDELS observations only cover the central part of ECDFS, i.e. the GOODS-South region.
The 12 bands images have distinct PSFs.
The \textit{HST} images have PSFs with FWHM $\sim 0\farcs1-0\farcs16$, compared to $\sim 1\arcsec$ for the MUSYC images and $\sim 0\farcs8$ for the CFHT images.\footnote{See \cite{An13} for a summary of these data.}

Instead of measuring fluxes within the same aperture from the 12 bands images downgraded to the worst PSF, we intend to maximize S/N for aperture-matched colors between the 12 bands.
We first determine colors for MUSYC, \textit{HST} and CFHT bands, respectively.
Second, we match three sets of colors to establish SEDs from the $U$ to the $K_{\rm s}$. 
The aperture-matched fluxes in the $U$, $B$, $V$, $R$ and $I$ (and thus colors between them) are available for 124 of the 140 targets from the MUSYC public catalog \citep{Cardamone10}.
The other 16 are optically too faint to be detected by MUSYC observations.  
For the \textit{HST} data set, $V_{606}$ and $z_{850}$ images are convolved with $H_{160}$ PSF, while $J_{125}$ and $H_{160}$ images are convolved with $z_{850}$ PSF.
These operations enable us to match the four images to the same spatial resolution.
An aperture of diameter of 1$\arcsec$ is used to measure fluxes from the convolved images and give aperture-matched colors between the four \textit{HST} bands.
Colors between CFHT $J$, $K_{\rm s}$ and $H_2S1$ bands are derived from photometry over the same area of a target using SExtractor in ``dual-image'' mode.
The next step is to match MUSYC, \textit{HST} and CFHT colors to the same aperture.
This is done by deriving the aperture-matched color between the $J_{125}$ ($z_{850}$ instead if $J_{125}$ is not available) and $J$ bands, and that between the $V_{606}$ and $V$ bands.
Since $J_{125}/z_{850}$-band PSF ($0\farcs13$/$0\farcs1$) is much smaller than $J$-band PSF ($0\farcs8$), we convolve the $J_{125}/z_{850}$ image with the $J$-band PSF to match the spatial resolution of $J$-band image.
An aperture of 2$\arcsec$ diameter is used to measure fluxes accordingly.
Moreover, the MUSYC colors are linked to the \textit{HST}-band colors by matching the interpolated $V_{606}$ between $V$- and $R$-band magnitudes with $V_{606}$.
The aperture-matched colors establish the $U-K_{\rm s}$ SED.
The SED is then scaled up to meet the total magnitude of $K_\mathrm{s}$ derived from aperture photometry within a diameter of 2$\arcsec$ corrected for missing flux out of the aperture (see \citealt{Retzlaff10} for more details). 
By doing so, we obtain SEDs for our 140 emission-line objects.

\begin{figure}
\centering
\includegraphics[width=0.45\textwidth]{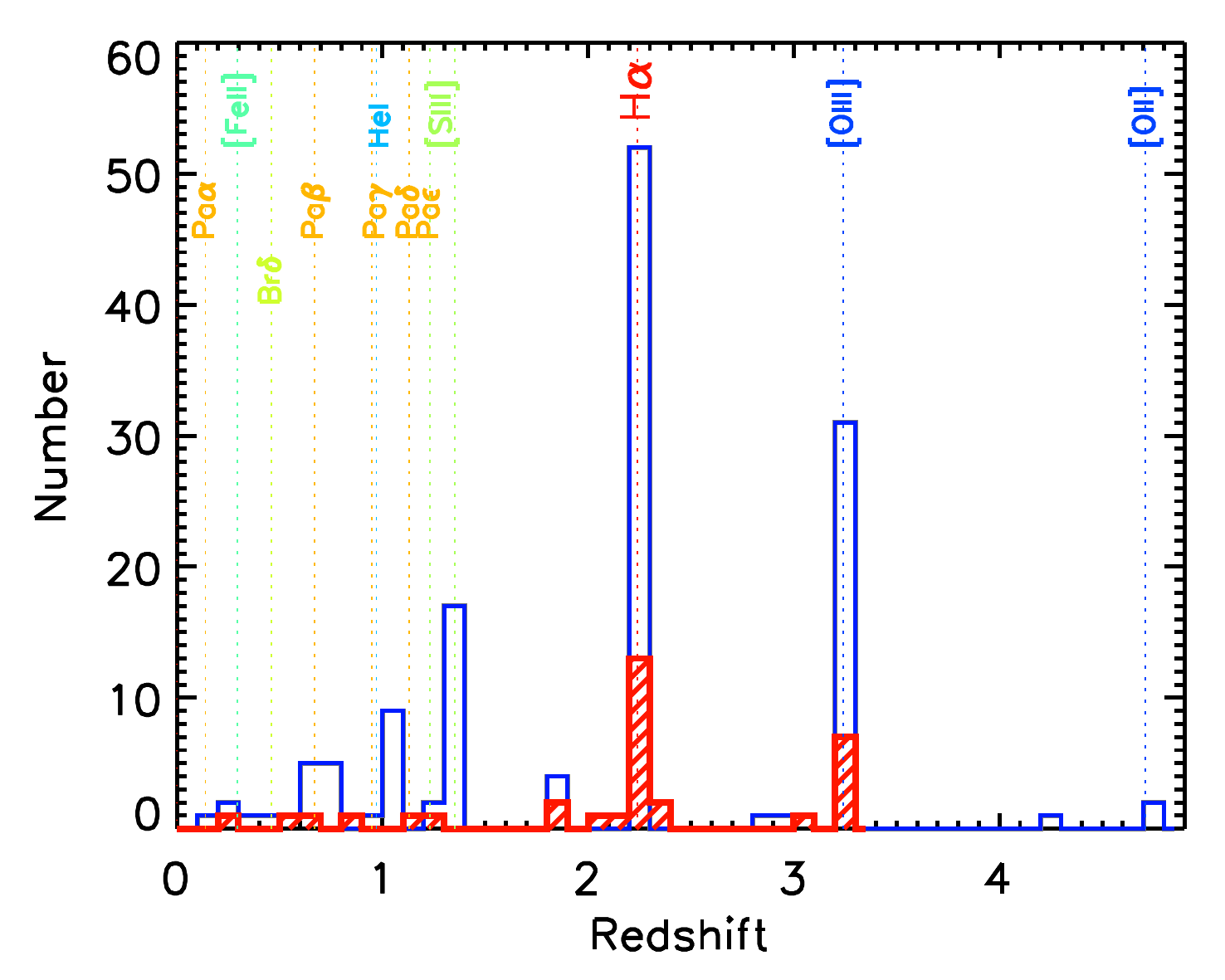}
\caption{Distribution of photo-$z$ for 138 selected emission-line candidates.
The red shaded area shows the distribution of 33 candidates with spec-$z$.
The dotted lines mark the emission-lines detected by the $H_\mathrm{2}S1$ filter at given redshifts.}
\label{f:cdfs_photz.eps}
\end{figure}

\begin{figure*}[!h]
\centering
\includegraphics[width=\textwidth]{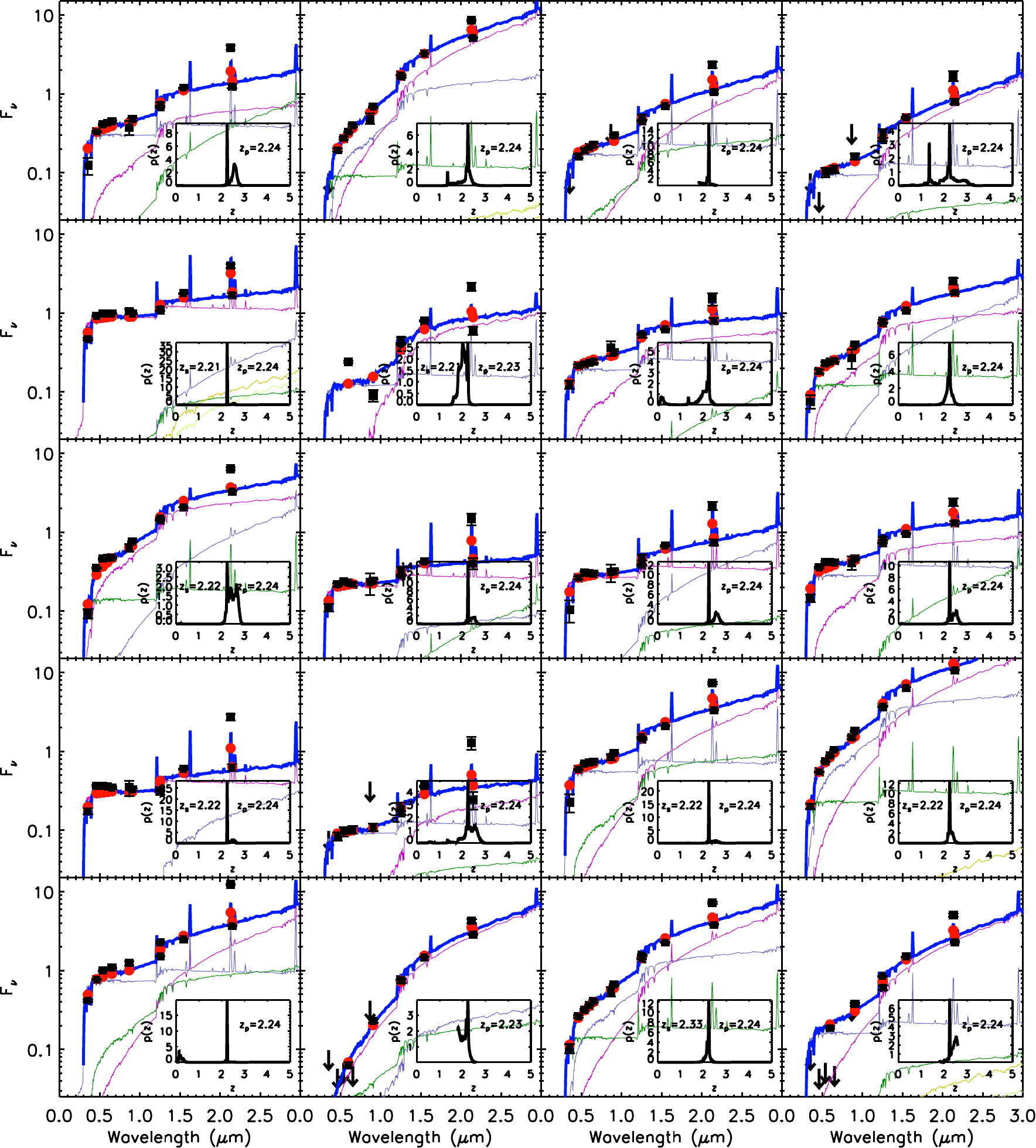}
\caption{Best-fit SEDs of \Ha\ ELGs. The squares are the observed data points and circles are the best-fit data points from EAZY code \citep{Brammer08}.
The arrows show the upper limits of corresponding bands, which means the source is not detected or resolved in these bands.
In each plan, the thick line shows the best-fit SED and light lines are the used templates; The inset plot shows the integrated probability of redshift.}
\label{f:sed.eps}
\end{figure*}

\begin{figure*}
\centering
\figurenum{3}
\includegraphics[width=\textwidth]{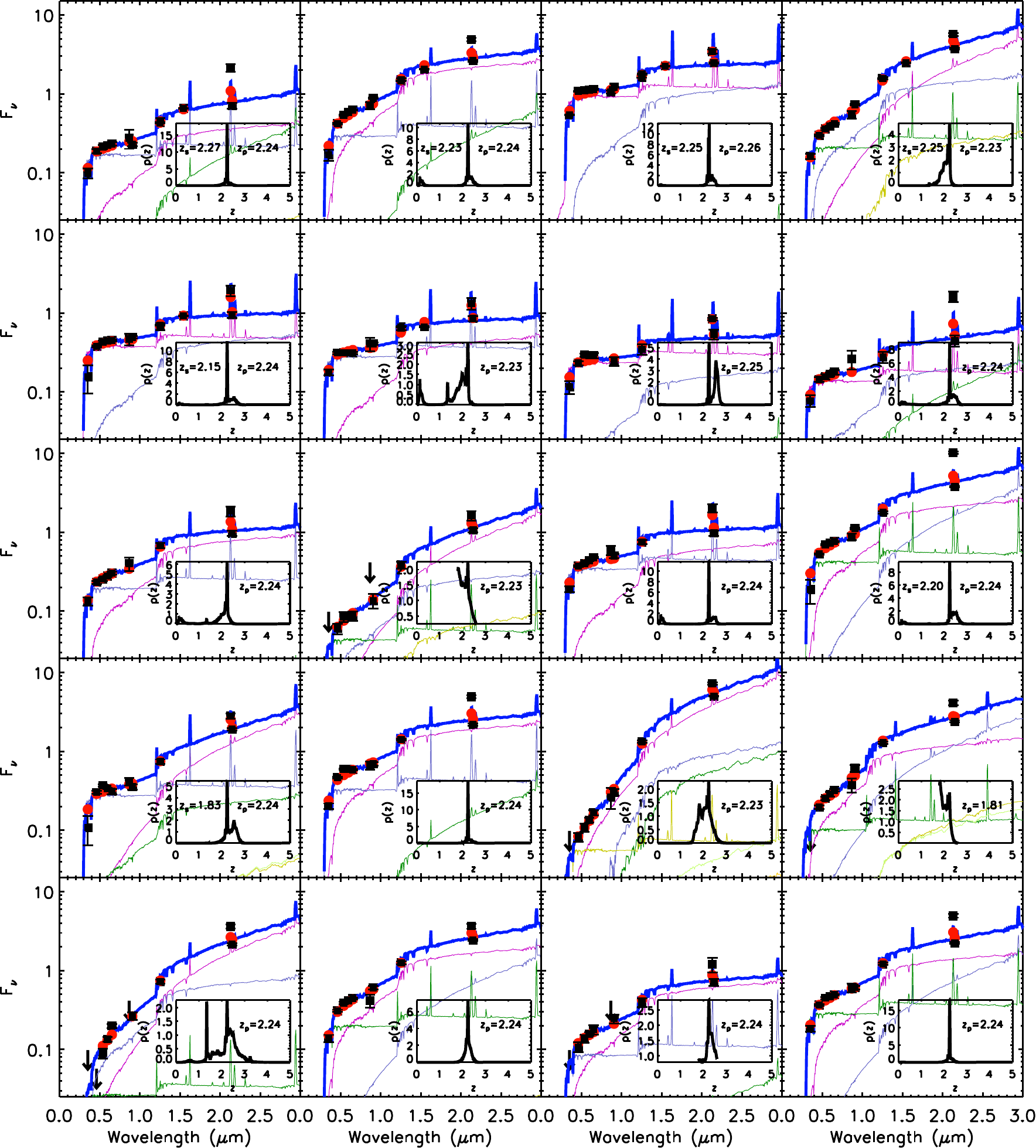}
\caption{(Continued).}
\end{figure*}

\begin{figure*}
\centering
\figurenum{3}
\includegraphics[width=\textwidth]{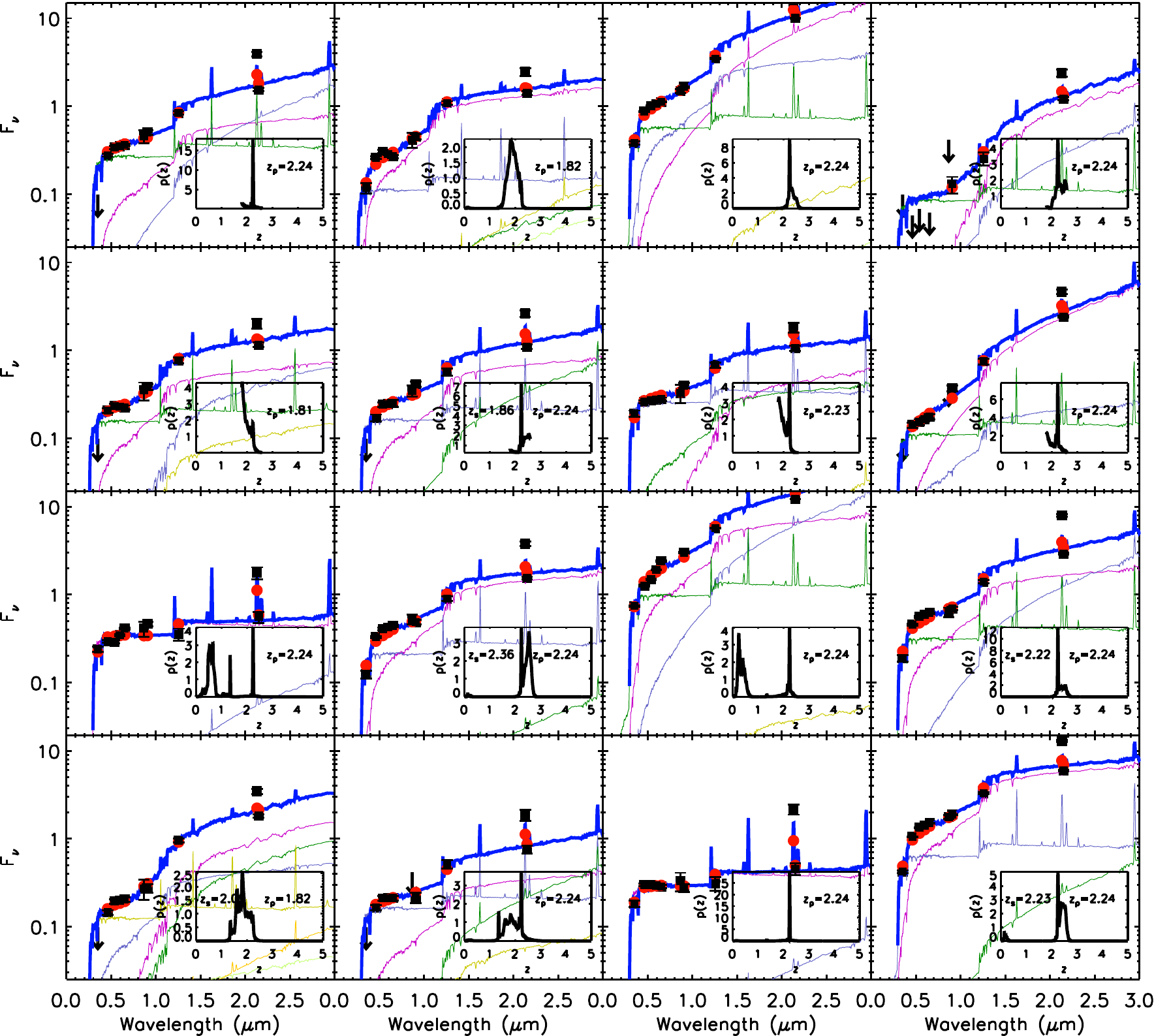}
\caption{(Continued).}
\end{figure*}

The software tool EAZY  \citep[Easy and Accurate Redshifts from Yale; ][]{Brammer08} is used to derive photometric redshifts (photo-$z$).
The template-fitting method used in EAZY to estimate photo-$z$ suffers from the fact that template color frequently degenerates with redshift. 
The $K_\mathrm{s}$ magnitude is then taken as a Bayesian prior to assigning a low probability to a very low redshift and a similarly low probability for finding extremely bright galaxies at high-$z$.
The emission-line objects selected with NIR narrow-band excess are mostly, if not all, SFGs \citep[e.g., see][]{Hayes10}. 
Therefore, the default library of galaxy templates in EAZY is chosen to derive photo-$z$. 
The library includes five templates generated based on the P{\'E}GASE population synthesis models \citep{Fioc97} and calibrated with synthetic photometry from semi-analytic models, and one template of young ($t=50$\,Myr) and dusty ($A_\mathrm{V}=2.75$) starbursts.
The combination of the six templates is able to model galaxies with colors over a broad range and minimize the color and redshift degeneracy.
Moreover, the P{\'E}GASE models provide a self-consistent treatment of emission lines.
Among 140 emission-line objects, two have SED composed of $<5$ valid data points and fail to produce photo-$z$. 

Figure~\ref{f:cdfs_photz.eps} shows the distribution of photo-$z$ for the 138 objects.
The 33 objects with spec-$z$ are shown by the hatched regions.
The emission lines potentially detected by $H_2S1$ are marked.
We can clearly see two prominent peaks at $z\sim 2.24$ and $z\sim 3.25$, corresponding to the emission lines \Ha\ and \OIII, respectively.
These two peaks are confirmed by the spec-$z$ distribution.
The good agreement between spec-$z$ and our photo-$z$ indicates that our $H_2S1$ data are critical to determining photo-$z$ to a relatively-high precision.
Other two peaks at $z\sim 1.3$ and $z\sim 1.0$ contain 17 and 10 objects, respectively.
The peak at $z\sim 1.3$ corresponds to \SIII$\lambda$9096, while the peak at $z\sim 1.0$ corresponds to P$\gamma\lambda$10935, P$\delta\lambda$10047 or He\,{\scriptsize I}$\lambda$10830 lines.
The relatively high abundance of \SIII$\lambda$9096 was seldom reported in the literature.
We further examine their properties in \cite{An13}.
We note that two objects have photo-$z$ corresponding to no known strong emission lines.
They are located close to the $\Sigma=3$ line in Figure~\ref{f:cdfs_mag_sele}.
We argue that the two objects are likely noise contaminators.

\input table1.tex

Accounting for potential uncertainties in photo-$z$ \citep[see also][]{Lee12}, we take 56 objects with $1.8<z_\mathrm{phot}<2.6$ as \Ha\ emitters at $z=2.24$.
Similarly, we identify 34 \OIII\ emitters at $z=3.25$, two \OII\ emitters at $z=6.72$, and other emitters at lower redshifts, including Pa$\alpha$ at $z=0.14$, \FeII\ at $z=0.30$ and \SIII\ at $z=1.23/1.35$.
Here the percentage of \Ha\ emitters among all emitters is 40\% (56/140), consistent with $\sim$36\% in \citet{Hayes10} and \citet{Lee12}.
We show the best-fit SED models and the probability function of photo-$z$ for the 56 \Ha\ emitters in Figure~\ref{f:sed.eps}.
Table~\ref{table1} lists their multi-band photometry and Table~\ref{table2} gives their photo-$z$ and spec-$z$.

\section{PROPERTIES OF \Ha\ EMISSION-LINE GALAXIES} \label{s:property}

\subsection{X-Ray Properties} \label{s:x-ray}

We first match our \Ha\ emitters with the X-ray source catalog of 4\,Ms \textit{Chandra} exposure in CDFS \citep{Xue11} to examine their X-ray properties.
Three out of 56 \Ha\ emitters are detected in X-ray (X-ray source ID = 29, 399, and 435 in the \textit{Chandra}  catalog), with absorption-corrected rest-frame 0.5-8\,keV luminosities $2.60\times 10^{43}, 2.76\times 10^{42}$, and $1.15\times 10^{43}$\,erg\,s$^{-1}$, respectively.
While the faintest X-ray among the three (X-ray source ID = 399) was identified as a galaxy, the other two more luminous X-ray sources were classified as active galactic nuclei (AGNs), both of which are likely obscured AGNs according to their X-ray hardness ratios \citep{Xue11}.
In addition, we find that the three \Ha\ sources have luminosity $L_\mathrm{H\alpha}>10^{43.5}$\,erg\,s$^{-1}$, being the most luminous ones in our sample and also the most massive ones with $M>10^{11}\,\Msun$ (see Section~\ref{s:mass}). 

For the other non-detected X-ray \Ha\ emitters, we perform X-ray stacking analyses using 4\,Ms \textit{Chandra} exposures.
We exclude sources with no X-ray coverage or with contaminations from nearby bright X-ray sources.
We finally stacked X-ray images of 42 \Ha\ emitters, yielding an effective exposure time of 106\,Ms.
We find a marginal detection in the stacked 0.5-2\,keV image (S/N = 2.5), and obtain an average rest frame 0.5-2.0\,keV X-ray luminosity of $3.5\pm1.4 \times 10^{42}$\,erg\,s$^{-1}$ by assuming a power-law spectrum with photon index of 2.0.
This confirms that the vast majority of our sample consists of star-forming ones.

\subsection{Morphologies} \label{s:morphology}

\textit{HST}/ACS $V_{606}$ and $z_{850}$ imaging data are used to examine morphologies.
The two bands correspond to the rest-frame 1839\AA\ and 2794\AA\ for $z=2.24$.
\textit{HST}/WFC3 imaging data from CANDELS are available for 26 of our 56 sample targets. 

\input table2.tex

\begin{figure*}
\centering
\includegraphics[width=0.93\textwidth]{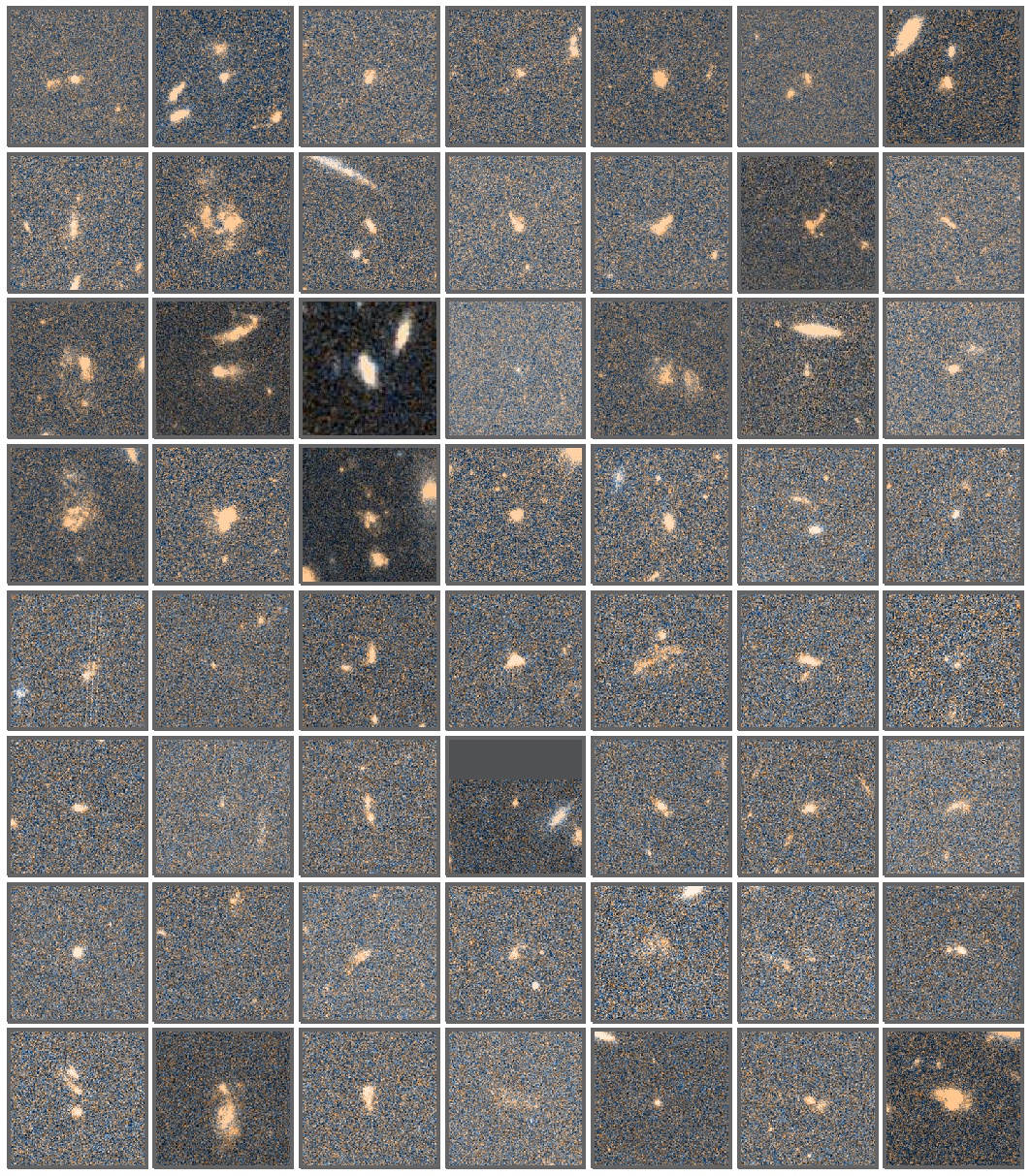}
\caption{The $6\arcsec \times 6\arcsec$ \textit{HST} color images are made of ACS $V_{606}$ and $z_{850}$ from GEMS and GOODS. 
The 56 \Ha\ ELGs listed in Table~\ref{table1} are given from left to right and top to bottom.
One image (the 3rd from the top and 3rd from the left) is made of \textit{HST}/WFC3 $J_{125}$ and $H_{160}$ from CANDELS due to no $V_{606}$ observation.} 
\label{f:H-alpha_F606W.eps}
\end{figure*}

\begin{figure*}
\centering
\includegraphics[width=0.93\textwidth]{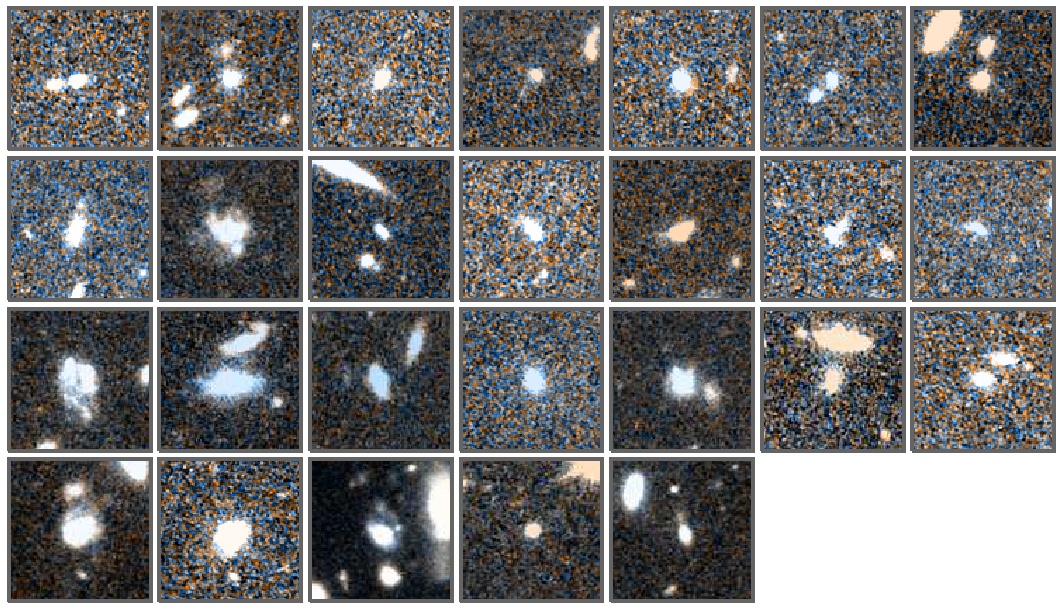}
\caption{\textit{HST}/WFC3 color images of the first 26 \Ha\ ELGs given in Table~\ref{table1}.
The color images are made of $J_{125}$ and $H_{160}$ from CANDELS.
Each thumbnail image has a size of $6\arcsec \times 6\arcsec$ corresponding to 49\,kpc$\times$49\,kpc at $z=2.24$.} 
\label{f:H-alpha_F125W.eps}
\end{figure*}

\begin{figure}
\centering
\includegraphics[width=0.45\textwidth]{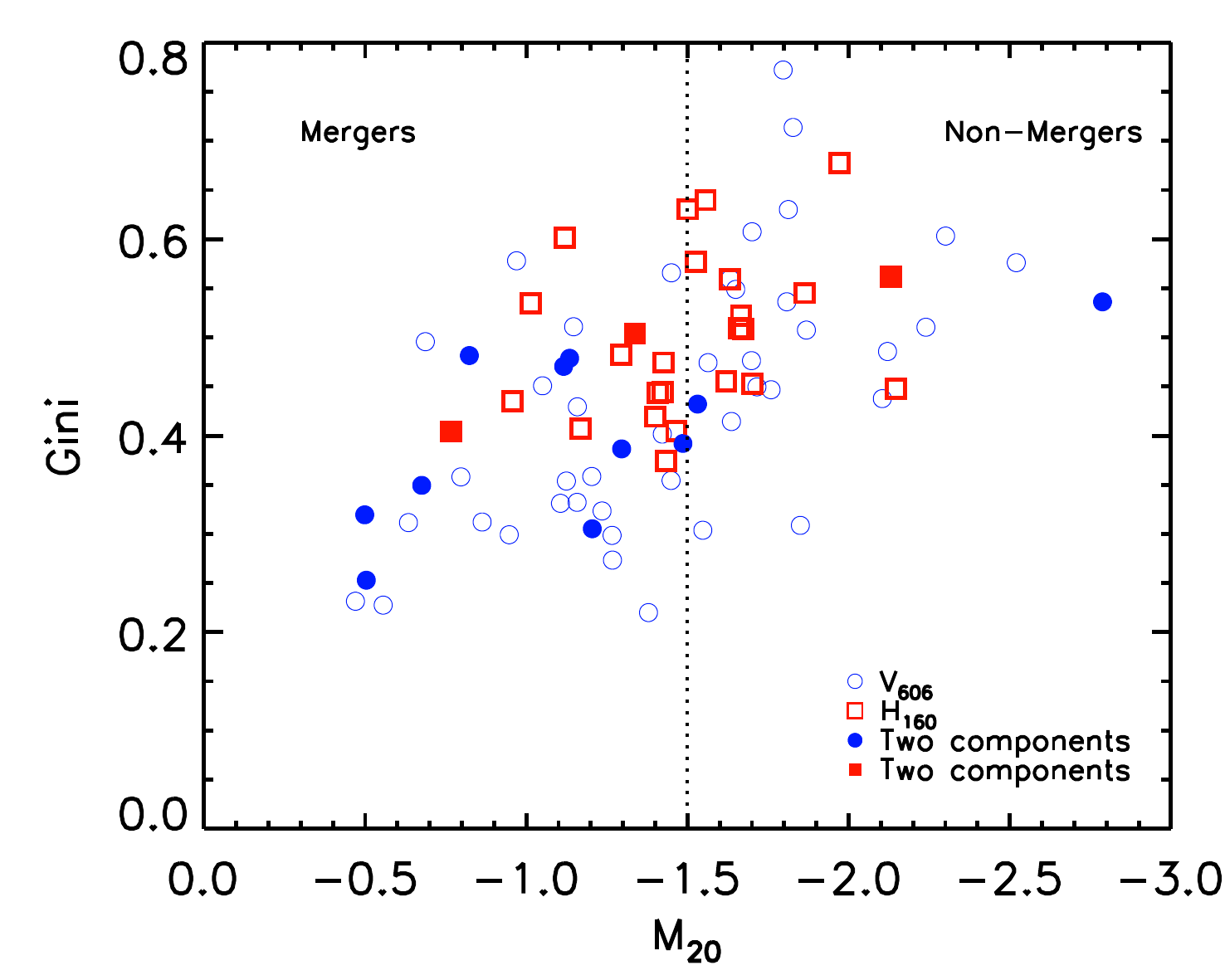}
\caption{ Gini vs. $M_\mathrm{20}$ coefficients for our sample of 56 \Ha\ ELGs.
The circles represent the results derived from the \textit{HST} F606W imaging data.
The squares are the results from \textit{HST} F160W imaging data for the 26 \Ha\ ELGs in GOODS-South.
The solid circles and solid squares are the sample galaxies with two major components.
The vertical line at $M_\mathrm{20} \sim -1.5$ is the empirical delineation between mergers and non-mergers \citep{Stott13}.} 
\label{f:H-alpha_mor_par.eps}
\end{figure}

Figure~\ref{f:H-alpha_F606W.eps} shows \textit{HST} $V_{606}$+$z_{850}$ color image stamps for 55 \Ha\ ELGs and $J_{125}$+$H_{160}$ color image stamp for one (the third from the top and third from the left). 
One can see that the \Ha\ ELGs in our sample show a variety of morphologies.
The 56 \Ha\ ELGs are visually classified by three of us (X.~Z.~Z, F.~X.~A and G.~W.~F) and the median types are given in Table~\ref{table2}.
About 20\% (11/56) of the sample galaxies contain two major components of similar color and are separated by $\sim 1\arcsec-2\arcsec$ ($8-17$\,kpc), which eliminates the possibility of them being multiple clumps within a single galaxy; 
25\% (14/56) appear to be merger remnants with tails, `tadpole', or a peculiar shape;
23\% (13/56) show spiral/diffuse/clumpy morphologies; 
18\% (10/56) are compact; 
14\% (8/56) are too faint to be resolved ($V_{606}<26$).
We point out that the $V_{606}/z_{850}$-faint \Ha\ galaxies are indeed dusty starbursts and their rest-frame UV lights are heavily attenuated.
The UV-selection (e.g., Lyman break technique) is unable to pick up such dusty SFGs.
This also suggests that the dust extinction of SFGs varies in a wide range.
Of three X-ray sources, two are found in merging systems and one is a compact galaxy.

We notice that roughly 10\% (6/56) of \Ha\ ELGs exhibit an apparent offset ($>0\farcs5$) between the central position in optical and NIR, indicating the complexity of morphologies in the rest-frame UV due to star formation activity and dust attenuation.
Figure~\ref{f:H-alpha_F125W.eps} shows \textit{HST} $J_{125}$+$H_{160}$ color image stamps for the first 26 of the 56 \Ha\ ELGs.
Still, our morphological classifications from $V_{606}$/$z_{850}$ images are mostly consistent with those from $J_{125}$/$H_{160}$ images for the 26 \Ha\ ELGs in GOODS-South.
The similarity in morphology between the rest-frame UV and the rest-frame optical for high-$z$ SFGs are also claimed by other studies \citep[e.g.,][]{Papovich05,Law12}.

We also measure the morphological parameters Gini (the relative distribution of the galaxy pixel flux values) and $M_\mathrm{20}$ \citep[the second-order moment of the brightest 20\% of the galaxy's flux coefficients; ][]{Lotz04} to quantify galaxy morphologies \citep{Fang12}.
Figure~\ref{f:H-alpha_mor_par.eps} shows the results.
About half of our sample has $M_\mathrm{20} > \sim -1.5$, satisfying the empirical cut for mergers \citep{Stott13}.
This is consistent with our result base on visual classification.

\subsection{Dust Extinction} \label{s:dust}

We estimate dust extinction from SEDs.
Generally speaking, dust extinction mostly effects the young stellar population of a galaxy because new stars are formed in dusty environments.
On the other hand, the young population can be divided into two components: unattenuated and attenuated.
As described in Section~\ref{s:Photometric Redshifts}, we made use of EAZY code to fit SEDs with models from combination of six representative galaxy templates.
One in the six templates represents a young and dusty starburst with $A_\mathrm{V}=2.75$ mag. 
The other five templates represent distinct stellar populations without dust attenuation.
Thus, the best-fit models to a galaxy SED implying not only the best photo-$z$, but also the percentage of each template in total \citep{Brammer08}.
Therefore, the young and dusty starburst template can be used to approximately trace the amount of dust attenuation in a galaxy.
The total effective extinction $A_\mathrm{V}$ to the entire galaxy relies on the fraction of the dusty starburst component in the SED.
Accordingly, we calculated $A_\mathrm{V}$ using the best-fit models composed of six templates given by EAZY.
Figure~\ref{f:sed.eps} shows the decomposition of the best-fit models into the six components.
Here the $A_\mathrm{V}$ is limited to $A_\mathrm{V} \le 2.75$ mag adopted by the dusty starburst template, although some \Ha-selected SFGs may be very dusty with $A({\rm H\alpha})$ equal to several magnitudes \citep[e.g.,][]{Garn10,Sobral12}.
We point out that this limitation is not crucial to our results and our main conclusions are not affected.
We come back to this point in Section~\ref{s:Results}.

In addition, 12 of the 56 \Ha\ ELGs are found to have 24\,\mm\ counterparts.
The 24\,\mm\ catalog is extracted from deep {\it Spitzer} 24\,\mm\ imaging of the Far-Infrared Deep Extragalactic Legacy  survey \citep{Dickinson07} using the PSF-fitting method of \cite{zheng06}.
The 3\,$\sigma$ detection limit is 24.6\,$\mu$\,Jy, corresponding to the 50\% completeness limit.
The cross correlation between the two catalogs is done with a matching radius of 1$\farcs$2.
We list their 24\,\mm\ magnitudes in Table~\ref{table1}.
We find that all the 12 objects have extinction $>$ 1\,mag and SFRs $>$ 20\,$\Msun\,{\rm yr}^{-1}$ (see Section~\ref{s:sfr}).
The 12 sample targets are indeed dusty starburst galaxies.
This confirms that our method works well in estimating dust extinction.

\subsection{\Ha\ Luminosities and SFRs}\label{s:sfr}

Following \citet{Ly11}, we derive \Ha\ + \NII\ flux from the narrow-band excess $K_\mathrm{s} - H_2S1$ and $K_\mathrm{s}$ total magnitude using the formula
\begin{equation}
  {F_{L} = \Delta{\rm NB} \frac{f_{\rm NB} - f_{K_{\rm s}}}{1-(\Delta{\rm NB}/\Delta K_{\rm s})}},
\end{equation}
where $f_{\rm NB}$ and $f_{K_{\rm s}}$ are fluxes given in  the units of erg\,s$^{-1}$\,cm$^{-2}$\,\AA$^{-1}$ for the $H_2S1$- and $K_{\rm s}$-band with filter width $\Delta{\rm NB}=293$\,\AA\ and $\Delta K_{\rm s}=3250$\,\AA, respectively.

We use \NII/\Ha\ = 0.117 to correct the contribution of \NII\,$\lambda\lambda$6548,6583 and obtain the observed \Ha\ fluxes \citep{Hayes10}.
The selection cut EW $>50$\,\AA\ together with the 5$\sigma$ depths of $H_2S1=22.8$ and $K_{\rm s}=24.8$ determines an \Ha\ flux detection limit $>2.1\times 10^{-17}$\,erg\,s$^{-1}$\,cm$^{-1}$\,Hz$^{-1}$.
Differing from low-$z$ SFGs, the high-$z$ ones seem to show little difference between extinction derived from continuum and nebular lines \citep{Erb06b}.
We employ the Calzetti extinction law to estimate attenuation to \Ha\ line $A({\rm H\alpha})$ from $A_\mathrm{V}$ which is derived from the SED modeling \citep{Calzetti00}.
The intrinsic \Ha\ fluxes are estimated from the observed \Ha\ fluxes with correction for extinction $A({\rm H\alpha})$.
Then \Ha\ luminosity is calculated with $z=2.24$.

The \Ha\ luminosity is an estimator of SFR.
We follow \citet[][hereafter K12]{Kennicutt12} to calculate SFR using 
\begin{equation} \label{e:sfr}
\log ({\rm SFR}/\Msun\,{\rm yr}^{-1}) = \log L_\mathrm{H\alpha} - 41.27,
\end{equation}
where $L_\mathrm{H\alpha}$ is the extinction-corrected \Ha\ luminosity in the units of erg\,s$^{-1}$.
The SFR is calibrated with STARBURST99 model \citep{Leitherer99} and Kroupa IMF, differing from that given in \citet[][hereafter K98]{Kennicutt98}.
Table~2 in \citet{Kennicutt12} gives the conversion between K12 and K98 calibrations.
For \Ha, SFR$_\mathrm{K12}/$SFR$_\mathrm{K98}=0.68$.

\begin{figure}
\centering
\includegraphics[width=0.45\textwidth]{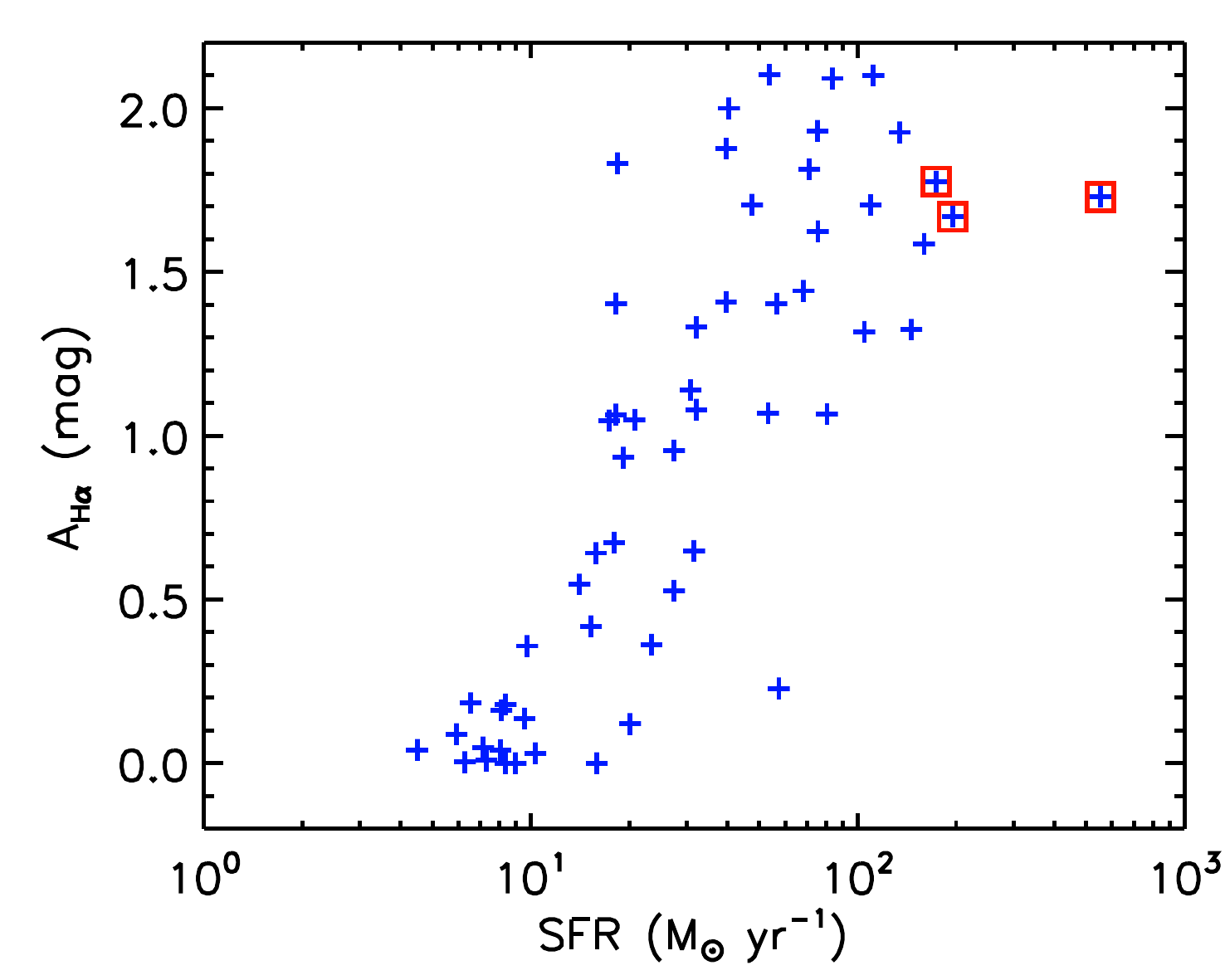}
\caption{Relationship between SFR and SED-derived extinction for our sample of 56 \Ha\ ELGs at $z$ = 2.24.
The red squares mark three X-ray sources.}
\label{f:atten_sfr.eps}
\end{figure}

Figure~\ref{f:atten_sfr.eps} shows the estimated SFR versus \Ha\ extinction.
The data points spread mainly from 3 to 300\,$\Msun\,{\rm yr}^{-1}$ in SFR and from 0 to 2.2\,mag in \Ha\ extinction $A({\rm H\alpha})$.
A correlation between SFR and H$\alpha$ extinction is apparent.
The increase of extinction at an increasing SFR found for \Ha\ ELGs is also seen for low-$z$ SFGs \citep[e.g.,][]{GarnBest10} and high-$z$ SFGs \citep{Ly12,Sobral12,Dominguez12,Kashino13}.

\subsection{Stellar Masses} \label{s:mass}

\begin{figure}
\centering
\includegraphics[width=0.45\textwidth]{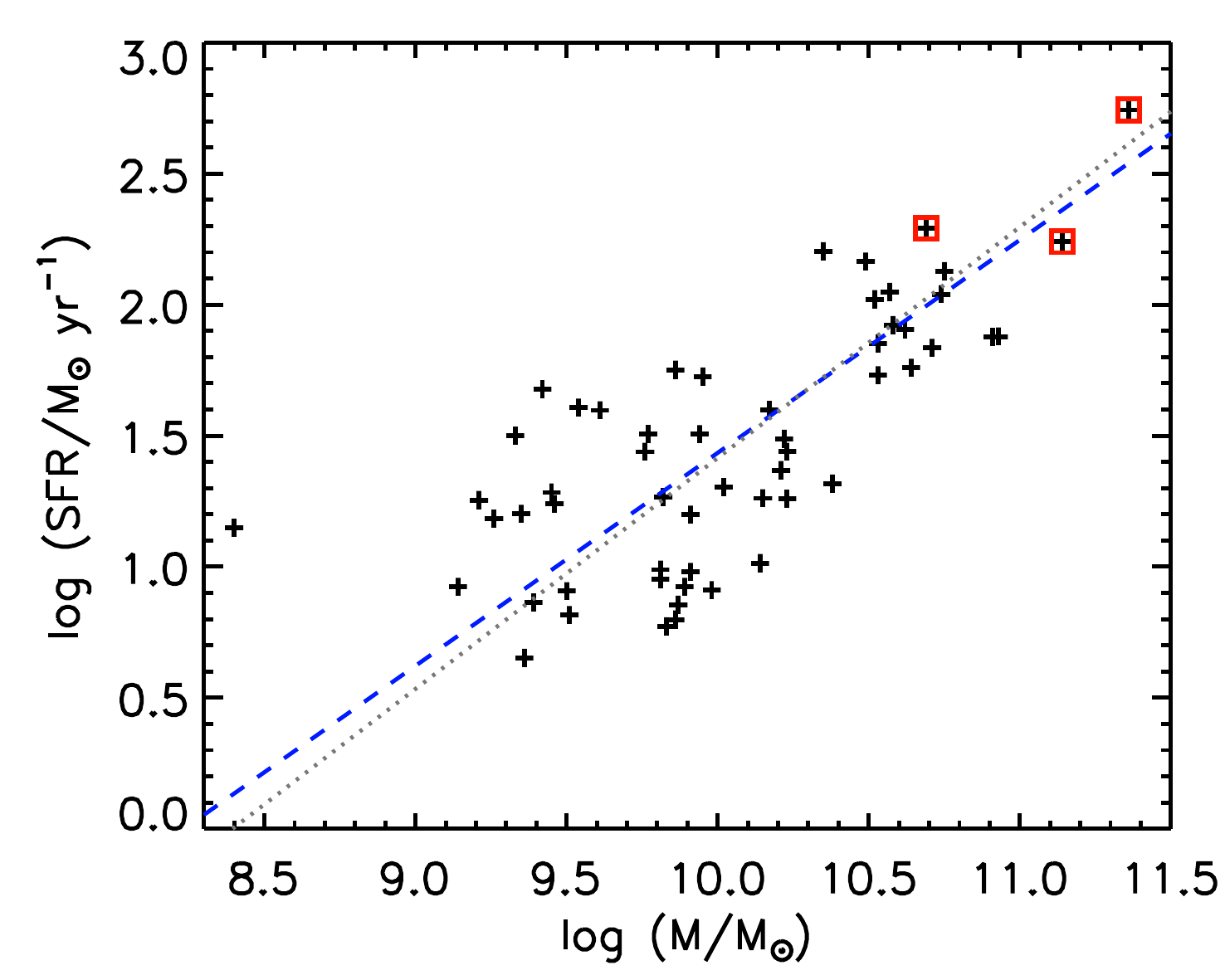}
\caption{The Mass-SFR relation of our 56 \Ha\ ELGs.
The three red squares are X-ray-detected AGNs.
The blue dashed line is a linear fit to the mass-SFR relation, giving a best-fit slope of 0.81.
The gray dotted line is the fit without the lowest stellar mass one, giving a best-fit slope of 0.88.
The three X-ray sources are removed in both fittings.}
\label{f:main_sequence.eps}
\end{figure}

The software FAST \citep[Fitting and Assessment of Synthetic Templates;][]{Kriek09} is used to estimate stellar mass for our sample.
The stellar population synthesis (SPS) models from \citet{Maraston05} and a Kroupa IMF \citep{Kroupa01} are adopted.
We fit the SPS models with exponentially declining star formation histories and the star formation timescale $\tau$ from 10\,Myr to 10\,Gyr in steps of 0.5\,dex.
The metallicity is fixed to solar ($Z$ = 0.02) and the dust extinction is modeled by the Calzetti reddening law, with $A_v$ from 0 to 3\,mag in steps of 0.01\,mag.
The ages of the model stellar populations range from 0.1\,Myr to the age of the universe.
The best photo-$z$ from EAZY is taken as input to FAST (see \citealt{Kriek09} for more details). 
Figure~\ref{f:main_sequence.eps} shows the distribution of our sample galaxies in stellar mass and SFR.
Three X-ray sources are marked with red squares.
The vast majority of the sample galaxies have a stellar mass $9<\log (M/\Msun)<11$.
The object has the lowest stellar mass $\log (M/\Msun)=8.40$ is faint in the $K_\mathrm{s}$-band but with a high EW and very blue color.
From Figure~\ref{f:main_sequence.eps}, we can see that the \Ha\ ELGs exhibit a correlation between stellar mass and SFR, which is known as the main sequence of SFGs \citep[e.g.,][]{Elbaz07,Daddi07,Wuyts11b}. 
The dashed line describes the best fit to the mass-SFR correlation (three X-ray sources are removed), having a slope of 0.81.
The scatter of this relation is 0.35\,dex.

\section{THE \Ha\ LF AT $z=2.24$} \label{s:Results}

One of our goals is to revisit the \Ha\ LF and examine the influence of extinction correction.
First, we commit Monte Carlo simulation to quantify the incompleteness of sample selection.
After that, we determine the function of intrinsic \Ha\ luminosities at $z=2.24$ using our sample of 53 X-ray-undetected \Ha\ galaxies.
Finally, we show that a constant extinction correction  to the observed \Ha\ luminosities results in an \Ha\ LF similar to the results in previous works.

\subsection{Completeness} \label{s:completeness}

A Monte Carlo simulation is used to derive the detection completeness as a function of intrinsic \Ha\ luminosity.
We assume that the intrinsic \Ha\ LF is described by a Schechter function with $L_\mathrm{H\alpha}^{*}=10^{42.88}$, $\alpha=-1.60$ and $\log \phi^{*}=-1.79$ from \citet{Sobral13}.
The $H_2S1$ filter centered at $\lambda_\mathrm{c}=2.130\,\micron$ has a width of $\Delta \lambda=0.0293$\,$\micron$, corresponding to $2.225<z<2.267$ for \Ha.
We use the averaged volume, corresponding to this redshift bin , as our effective volume.
The brighter ELGs are detectable over a wider range of the filter transmission than the faint one.
Therefore, we adopt a wider redshift span of $2.2<z<2.29$ to account for the entire wavelength coverage of the filter curve in our simulation.
Then we can estimate the contamination rate of bright ELGs from low-$z$ (2.200 $< z <$ 2.225) and high-$z$ (2.267 $< z< $ 2.290).
We generated a mock catalog of 30\,million galaxies with \Ha\ emission lines satisfying the given LF and the given redshift range.
In practice, the simulated galaxies are divided into 30 redshift bins.
In each redshift bin, the galaxies are divided again into 500 bins between $40<\log (L_\mathrm{H\alpha}/{\rm erg\,s^{-1}})<50$.
Each bin contains on average 2000 galaxies.
The contribution of \NII\ is added to the simulated \Ha\ using \NII/\Ha\ = 0.117.
Dust attenuation is applied following the relation given in Figure~\ref{f:atten_sfr.eps} with a scatter of 0.38 in $A({\rm H\alpha})$ taken into account.
Furthermore, a Gaussian line profile with $\sigma=200$\,km\,s$^{-1}$ is adopted to convolve \Ha\ at given redshifts with the $H_2S1$ filter transmission curve to simulate the observed \Ha+\NII\ fluxes.

\begin{figure}
\centering
\includegraphics[width=0.45\textwidth]{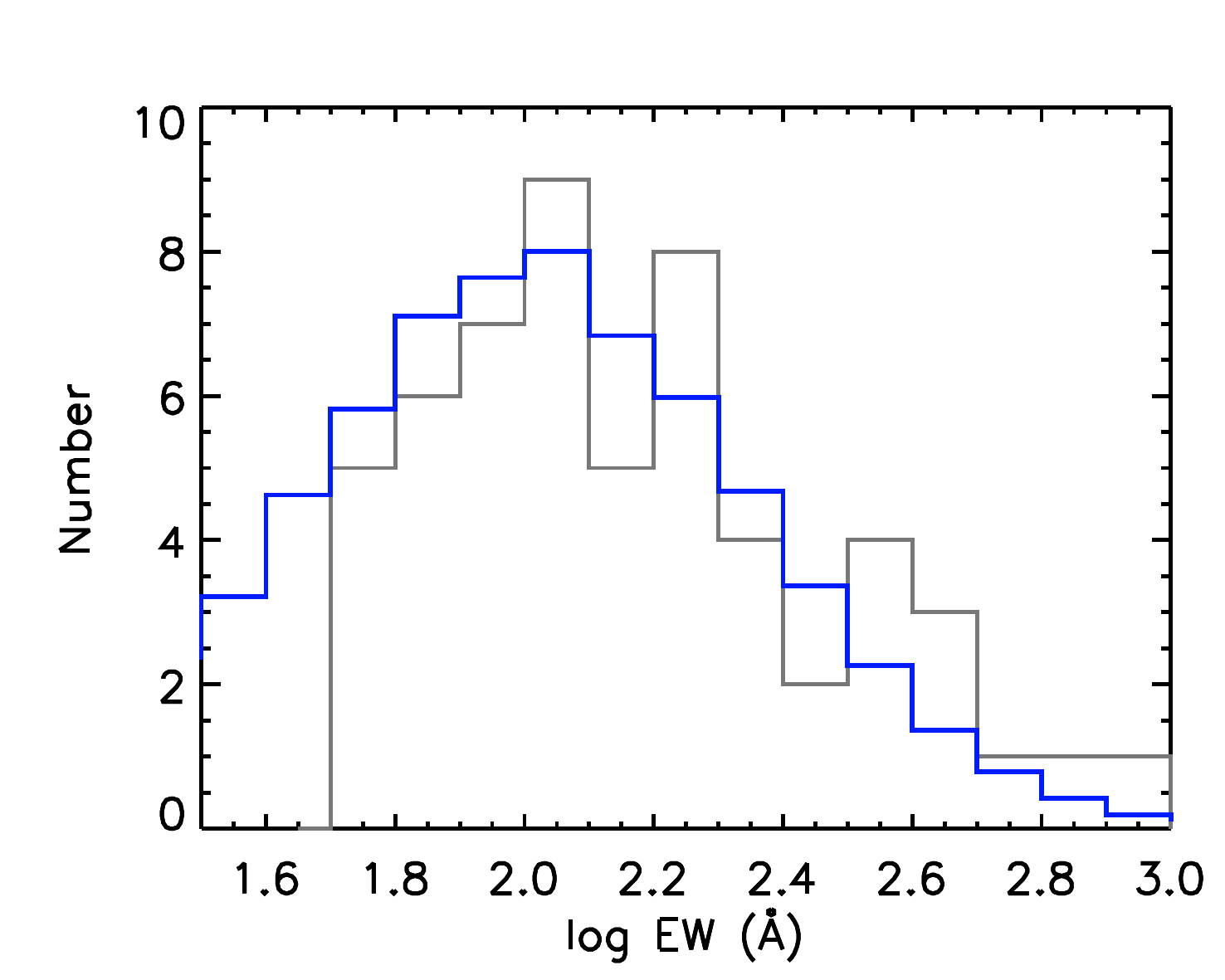}
\caption{Comparison of the modeled and observed EW distributions.
The blue thick line shows the best matched model with mean $\log ($EW$_{\rm rest})$ = 2.00 and $\sigma \,[\log ($EW$\mathrm{rest}/$\AA)] = 0.35 and is scaled to match the observed distribution (gray line).}
\label{f:ew.eps}
\end{figure}

To link \Ha\ + \NII\ of a given flux to $H_2S1$ and $K_\mathrm{s}$ magnitudes of a galaxy, one needs to know the EW of the line.
The intrinsic distribution of EW of \Ha\ + \NII\ is fairly similar to the observed distribution of rest-frame EW \citep{Lee07}.
We assume a log-normal distribution of EW for \Ha\ + \NII\ following \citet{Ly11}.   
Since the observed EWs of our sample are affected by noise as shown in Figure~\ref{f:cdfs_mag_sele},  we take the noise effects into account and infer the best-matched distribution as follows. 
A set of log-normal EW distributions are generated with the mean $\log ($EW$_{\rm rest})$ spanning between 1.8 and 2.3 and the spread $\sigma \,[\log ($EW$_{\rm rest}$/\AA)] between 0.15 and 0.65.
A step of 0.1\,dex is adopted for the two parameters.
Assuming that the flux of \Ha\ + \NII\ is not correlated with its EW, we randomly assign EWs satisfying a modeled EW distribution to the observed \Ha\ + \NII\ fluxes and give $H_2S1$ and $K_\mathrm{s}$ magnitudes.
Added to the photon noise and sky background noise from our $H_2S1$ and $K_\mathrm{s}$ images \citep[see also][]{Ly12}, the simulated galaxies can be selected by $H_2S1-K_\mathrm{s}$ selection criteria as shown in Figure~\ref{f:cdfs_mag_sele}.
By doing so, we model the `observed' EW distribution for each of input intrinsic EW distributions under the same conditions as our $H_2S1$ and $K_{\rm s}$ observations.
Comparing the modeled EW distributions with the observed EW distribution of our sample and minimizing the $\chi^{2}$, we obtain the intrinsic EW distribution best matching the observed EW distribution. 
The model with mean $\log ($EW$_{\rm rest})$ = 2.00 and $\sigma \,[\log ($EW$_{\rm rest}/$\AA)] = 0.35 best reproduces the observed distribution of our sample galaxies.
We show the best-matched mode in comparison with the observed distribution in Figure~\ref{f:ew.eps}.
This distribution is taken to randomly assign EWs to the mock catalog of simulated \Ha\ + \NII\ galaxies, thus making it possible to obtain $H_2S1$ and $K_\mathrm{s}$ magnitudes.

\begin{figure}
\centering
\includegraphics[width=0.45\textwidth]{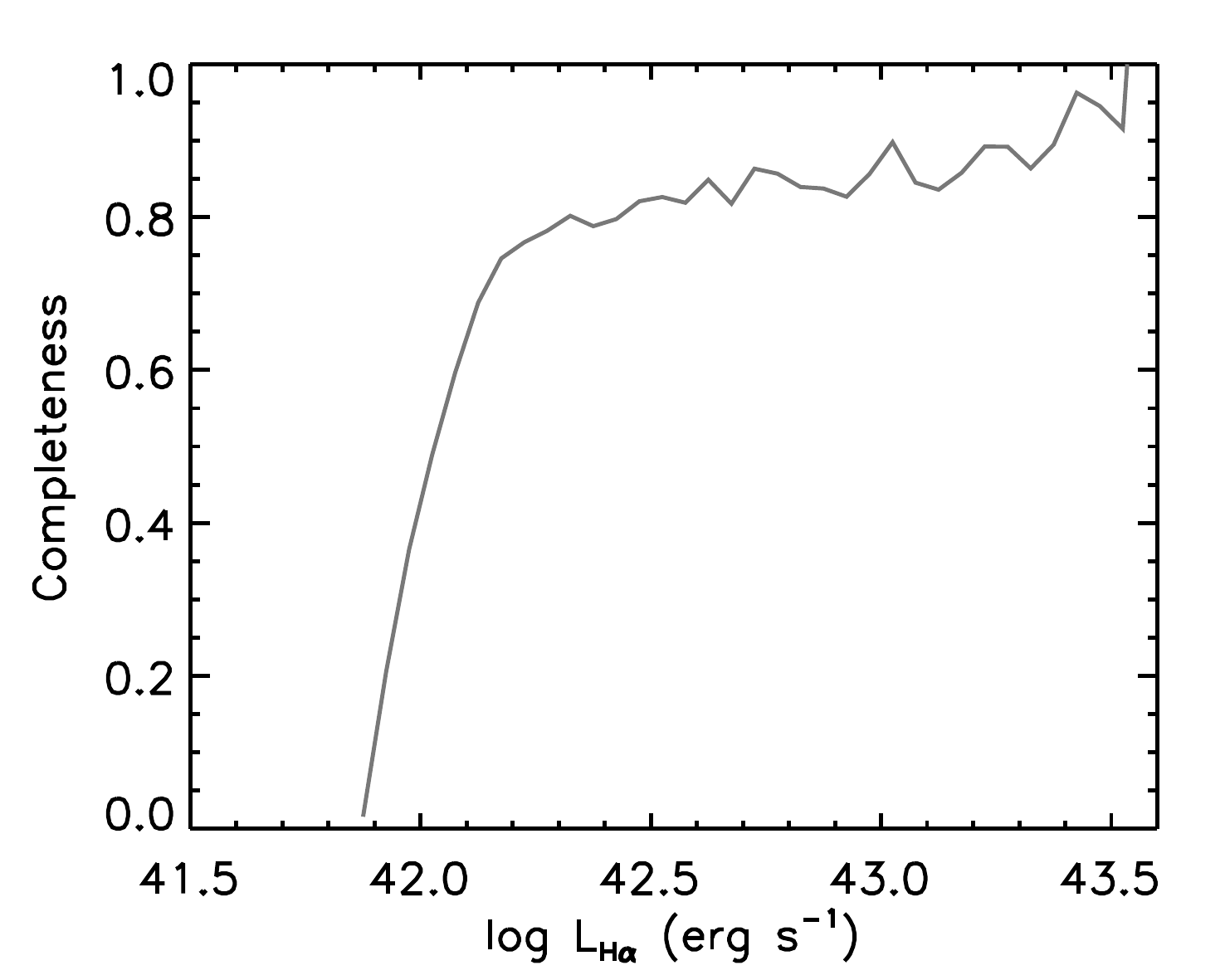}
\caption{Completeness as a function of \Ha\ luminosity.}
\label{f:comples.eps}
\end{figure}

Accounting for the photon noise and sky noises from our $H_2S1$ and $K_{\rm s}$ images, the simulated \Ha\ ELGs are selected using the criteria as shown in Figure~\ref{f:cdfs_mag_sele}.
By doing so, we obtained the fraction of the mock galaxies picked up by our selection at given intrinsic \Ha\ fluxes, giving the completeness as a function of the intrinsic \Ha\ luminosity shown in Figure~\ref{f:comples.eps}. 
Because the completeness curve is insensitive to the input Schechter function in the simulation, we employ the completeness curve given in Figure~\ref{f:comples.eps} to determine \Ha\ LF using our sample.
Note that the volume correction and completeness estimate for our sample root on the redshift bin $2.225<z<2.267$ and the final completeness curve accounts for all major effects involved in our observation and selection.

\subsection{Determination of \Ha\ LF} \label{s:Luminosity Function}

\begin{figure}
\centering
\includegraphics[width=0.45\textwidth]{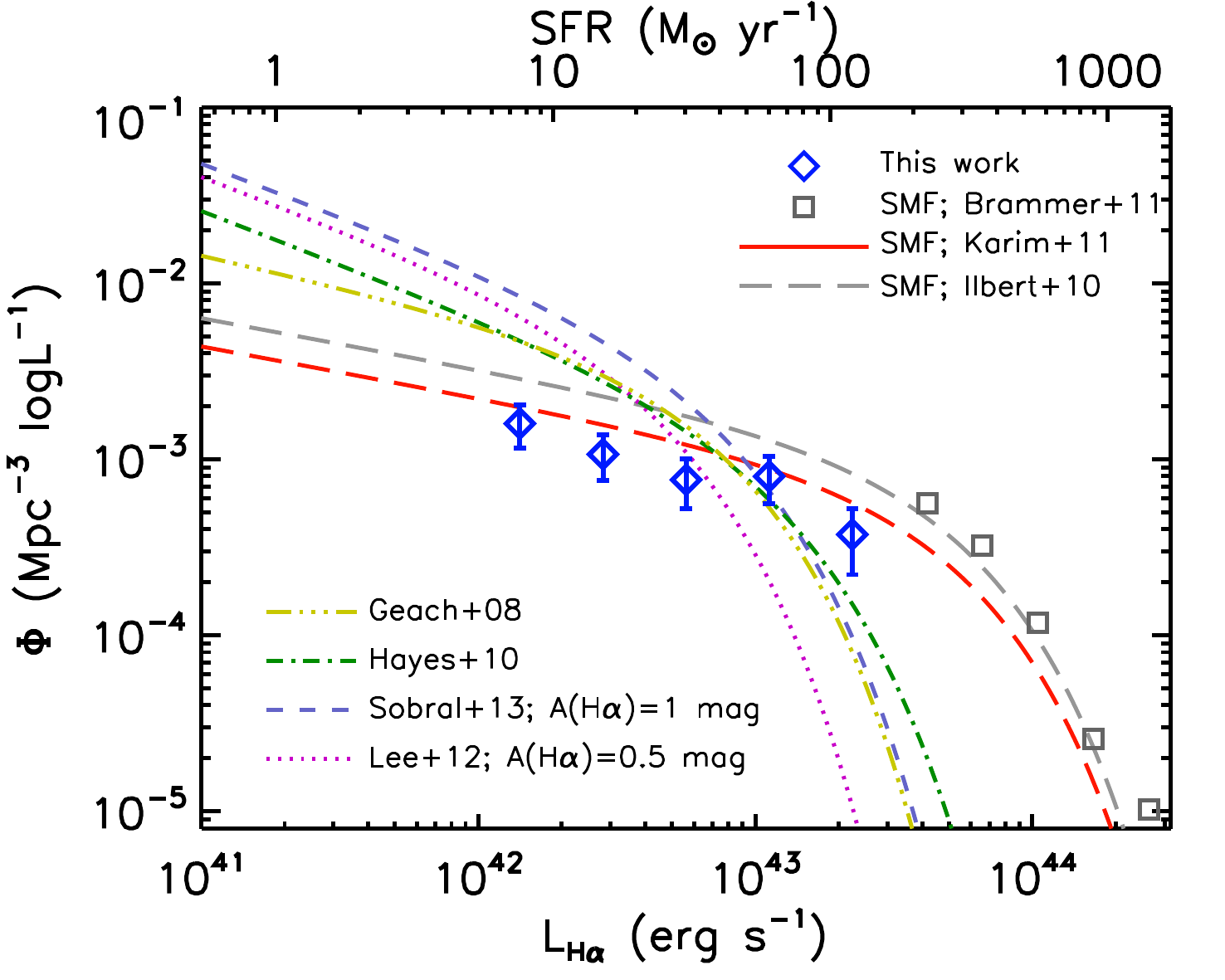}
\caption{\Ha\ LF at $z$ = 2.24.
The blue diamonds are data points based on individual extinction correction derived from SED modeling. 
The error bars are determined by Poisson noise.
The dashed lines are the SMF of SFGs at $1.5<z<2.0$ from \citet{Ilbert10} and that at $2.0<z<2.5$ from \citet{Karim11}.
The gray squares are the SMF of SFGs at $1.8<z<2.2$ from \citet{Brammer11}.
The thin lines show representative \Ha\ LFs at $z=2.23$ in the literature.}
\label{f:H_alpha_lf_sfr.eps}
\end{figure}

We determine the \Ha\ LF at $z=2.24$ using our sample of 53 \Ha\ ELGs.
Three X-ray sources are excluded to avoid AGN contamination.
A volume of $5.40\times 10^4$\,Mpc$^3$ is calculated from the coverage of 383 arcmin$^2$ of our sample and a redshift span $2.225<z<2.267$.
By dividing our sample into five bins in $\log (L_{\rm H\alpha})$ from 42.0 to 43.5 and taking the completeness correction, we obtained our \Ha\ LF data points as shown in Figure~\ref{f:H_alpha_lf_sfr.eps}.
The error bars represent mainly the Poisson noise.
We fit the data points with a Schechter function \citep{Schechter76} in the form of \\
\begin{equation} \label{e:schechter}
\begin{split}
&\Phi (\log L) \,d(\log L) =\\
&\ln(10)\, \phistar \,10^{(\alpha +1)(\log L- \log \Lstar)}\, \exp[-10^{\log L -\log \Lstar}]\, d(\log L),
\end{split}
\end{equation}
where $\Lstar$ is the characteristic luminosity, $\phistar$ is the normalization, and $\alpha$ is the faint-end power-law index.
The standard $\chi^{2}$ minimizer is used in the fitting and the errors are the formal 1$\sigma$ statistical errors on the parameters.
Since all our data points place in the power-law region of the LF, only the faint-end slope can be reliably constrained.
The best-fit faint-end slope is $\alpha=-1.36\pm0.32$.

It is worth noting that \Ha\ LF traces SFR function.
With the main sequence of SFGs, i.e., the tight correlation between SFR and stellar mass, one would expect that the \Ha\ LF is also connected with SMF.
We use the mass-SFR relation of SFGs at $z\sim 2$ (slope = 1 and $\log ($SFR$/\Msun$\,yr$^{-1}) =2.45$ at $\log(M/\Msun)=11.0$) from \citet[][]{Wuyts11b} to convert an SMF to an SFR function, and use Equation~\ref{e:sfr} to translate the SFR function into an \Ha\ LF.
Three SMFs of SFGs are adopted for comparison: $1.5<z<2.0$ from \citet{Ilbert10}, $2.0<z<2.5$ from \citet{Karim11} and $1.8<z<2.2$ from \citet{Brammer11}.
The \Ha\ LFs converted from the SMFs are plotted in Figure~\ref{f:H_alpha_lf_sfr.eps}.
Apparently, these \Ha\ LFs are in very good agreement with our \Ha\ LF in shape, although they tend to be slightly higher.
The stellar mass at the ``knee'' of the SMF is $\Mstar = 10^{10.95\pm0.03}$\,${\Msun}$ in \citet{Karim11}, corresponding $\Lstar = 10^{43.67}$\,erg\,s$^{-1}$ for \Ha\ LF.
Taking this value, we refit \Ha\ LF with the Schechter function and also with three X-ray sources removed, giving the best-fit result $\log\phistar=-3.27^{+0.14}_{-0.22}$ and $\alpha=-1.31\pm0.16$.
In summary, all of our two fittings of \Ha\ LF give a consistent faint-end slope.
Therefore, we conclude that the faint-end slope with the value of $\alpha = -1.3$ for \Ha\ LF is robust, despite our sample  not being large, and provides poor constraint on the bright-end of \Ha\ LF.
Therefore, the limitation of $A_\mathrm{V} \le 2.75$\,mag in estimating dust extinction as we described in Section~\ref{s:dust} does not affect our main conclusions.

The \Ha\ LFs at $z=2.23$ from previous works are also shown in Figure~\ref{f:H_alpha_lf_sfr.eps} for comparison.
Note that a constant extinction correction was adopted in these studies ($A({\rm H\alpha})=1$\,mag: \citealt{Geach08}, \citealt{Hayes10}, \citealt{Sobral13}; $A({\rm H\alpha})=0.5$\,mag: \citealt{Lee12}).
Large discrepancies can be seen between our \Ha\ LF and others: ours is much shallower at the faint end and higher at the bright end. 
We argue that the discrepancies are mostly attributed to the difference in extinction correction.
A fixed correction does not change the shape of an observed \Ha\ LF.
Our correction for extinction on an individual basis is able to recover heavily attenuated SFGs with intrinsically luminous \Ha\ from the faint end of the \Ha\ LF.
This enables the data points to spread into a wider luminosity range, resulting a decrease at the faint end and an increase at the bright end.

\begin{figure}
\centering
\includegraphics[width=0.45\textwidth]{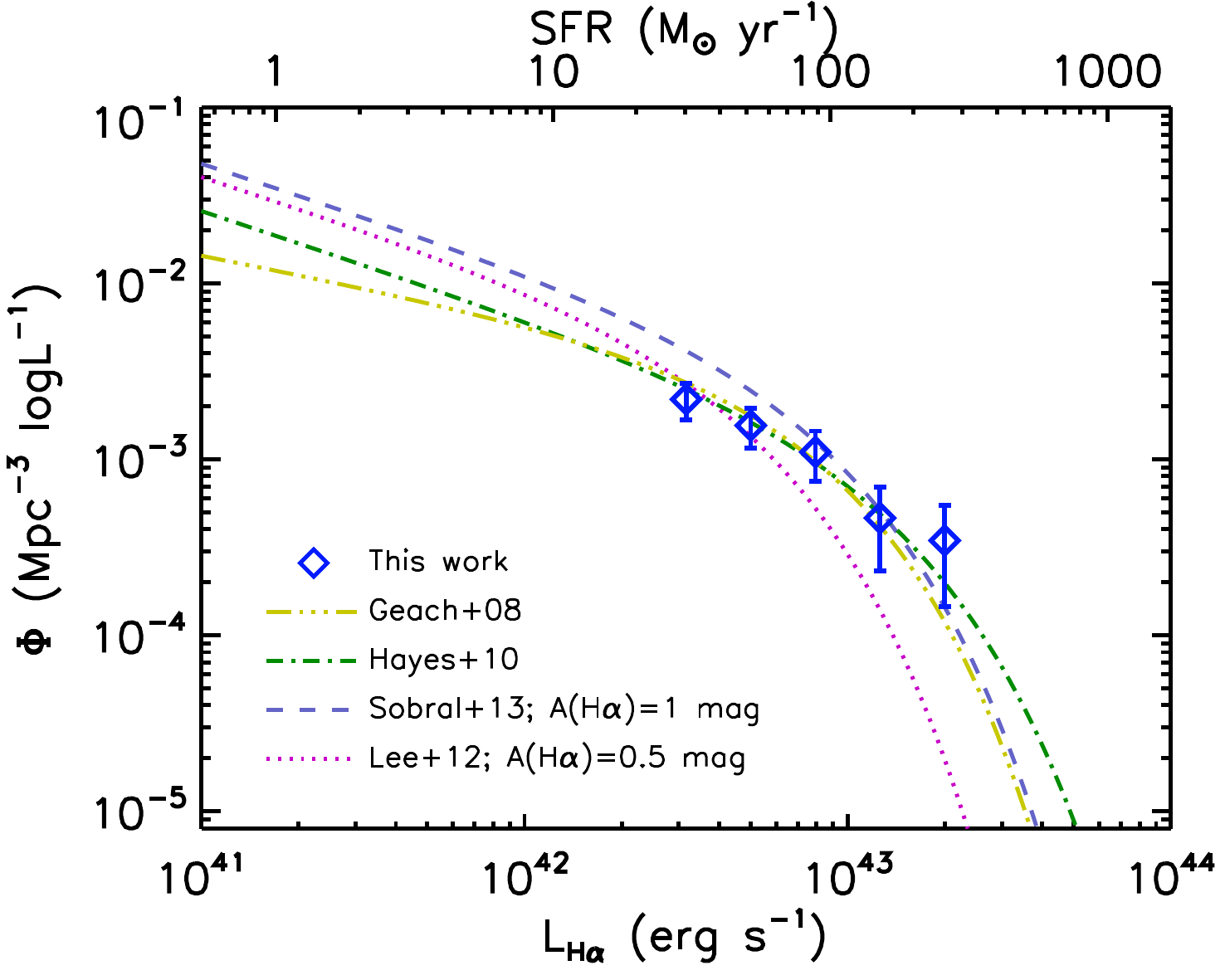}
\caption{\Ha\ LF derived with constant extinction correction.
The blue diamonds are the data points from this work.
The thin lines show the \Ha\ LFs from previous works with constant extinction correction: $A({\rm H\alpha})=0.5$\,mag in \cite{Lee12} and $A({\rm H\alpha})=1$\,mag in other works.}
\label{f:H_alpha_lf_sfr_const.eps}
\end{figure}

We rebuild an \Ha\ LF based on our sample but with a constant extinction correction ($A({\rm H\alpha})=1$\,mag).
Simulation is carried out with the constant extinction to derive the completeness as a function of \Ha\ luminosity.
We present the rebuilt \Ha\ LF in Figure~\ref{f:H_alpha_lf_sfr_const.eps}.
We also fit it with a Schechter function, giving a best-fit faint-end slope $\alpha=-1.59\pm0.26$.
This is consistent with the typical value of $\alpha \sim -1.6$ given in other works.
The same as in Figure~\ref{f:H_alpha_lf_sfr.eps}, the \Ha\ LFs from previous studies are shown for comparison.
A good agreement between these \Ha\ LFs is seen within uncertainties.
It becomes clear that extinction correction is key in determining the \Ha\ LF.
Furthermore, the extinction correction must be done on an individual basis.

\section{DISCUSSION AND CONCLUSION} \label{s:discussion}

We have described a search for \Ha\ ELGs using deep $H_2S1$ imaging combined with deep $K_\mathrm{s}$ imaging data of ECDFS. 
In total 140 emission-line candidates are identified. The deep optical and NIR imaging data in the ECDFS are utilized to construct broad-band SEDs and derive photo-$z$, stellar mass and extinction. 
We identify 56 \Ha\ emitters with \Ha\ flux $>2.1\times 10^{-17}$\,erg\,s$^{-1}$\,cm$^{-1}$\,Hz$^{-1}$. 
Three of the 56 are X-ray sources detected in the \textit{Chandra} 4\,Ms observation and two of them host an AGN in terms of their X-ray luminosities. 
The stacked X-ray luminosity of the rest is consistent with that of SFGs.
In our sample, only $\sim$ 4\% are contaminated by AGN, although obscured AGNs individually undetected in the current X-ray observation may still present in some of the sample galaxies. 
However, the obscured AGNs are unlikely to be responsible for exciting \Ha\ line.
As a result, our sample is clean for studying the properties of \Ha\ ELGs and determining the \Ha\ LF. 

\textit{HST}/ACS imaging reveals that nearly half of our \Ha\ ELGs are either mergers or close pairs. 
A similar merger fraction was suggested by \citet{Conselice03} for SFGs at $z\sim 2$, although large uncertainties exist in measuring merger fraction as a function of redshift \citep[e.g., $\sim$30\% for \Ha\ ELGs at $z=2.23$ by][]{Stott13}. 
A significant fraction (14\%) of the \Ha\ ELGs are too faint to be resolved in the rest-frame UV. 
These UV-faint ones are confirmed to be heavily-obscured SFGs by SED modeling.
The remaining one third exhibit extended or compact morphologies.  
We note that the morphological properties of our \Ha\ ELGs are consistent with those of high-$z$ SFGs \citep[e.g.,][]{Lotz06,Ravindranath06,Conselice11}. 
Our results reveal that merging/interacting processes play a key role in regulating star formation in typical SFGs at cosmic epoch $z\sim 2-3$. 

Our sample of 56 \Ha\ ELGs are representative of SFGs at $z\sim 2-3$ in many aspects. 
SFGs are found to be more obscured at higher SFR \citep[e.g.,][for $z\sim 0$; \citealt{Zheng07,Ly12,Dominguez12} for high-$z$]{Bothwell11}. 
We also find such a tendency among \Ha\ ELGs as shown in Figure~\ref{f:atten_sfr.eps}. Our extinction is estimated from SEDs. 
It is important to note that the SED-derived extinction tends to be lower than the true case represented by the IR to UV luminosity ratio $L_\mathrm{IR}/L_\mathrm{UV}$, especially at the high-SFR end \citep{Wuyts11a}.
This is understandable because in starbursts, parts of newly formed stars are deeply buried in dusty molecular clouds and can only be traced by the IR emission \citep[e.g.,][]{Swinbank04,Sedgwick13}.
Adding this correction would enhance the correlation between extinction and SFR. 
The mass-SFR relation is well-established from the present day all the way to $z\sim 3$ \citep[e.g.,][]{Elbaz07,Daddi07}. 
The slope of the mass-SFR relation at $z\sim 2$ is close to unity \citep{Wuyts11b}.
As shown in Figure~\ref{f:main_sequence.eps}, \Ha\ ELGs follow the mass-SFR relation of SFGs, giving a slope of 0.81.
We argue that the shallower slope of the \Ha\ ELGs is likely attributed to the underestimate of extinction at high end. 

The correction for extinction on an individual basis is vital to determining the shape of the \Ha\ LF. 
With extinction correction for individual sample galaxies, we build the \Ha\ LF at $z=2.24$. 
The completeness as a function of \Ha\ luminosity is carefully estimated via Monte Carlo simulation and is applied to the observed LF. 
Our sample is not large and unable to give a good constraint on the bright-end of \Ha\ LF.
A shallow faint-end slope of $\alpha=-1.32\pm 0.24$ is robust obtained, compared to the typical value of $\sim -1.6$ given by previous studies in which correction for a constant extinction is often adopted \citep{Geach08,Hayes10,Sobral13}.  
We have shown that a steep faint-end slope is inevitable when a fixed extinction is applied.
\citet{Garn10} computed the \Ha\ LFs at $z=0.84$ with correction for SFR-dependent extinction and constant ($A({\rm H\alpha})=1$\,mag) extinction, respectively, finding that the former gives a shallower faint-end slope. 
It is obvious that the constant extinction correction is not a good approximation because it will overestimate the extinction in faint \Ha\ ELGs and severely underestimate the extinction for brightest ones.
Resolving heavily-attenuated SFGs from the emitters with faint \Ha\ naturally leads to a decline at the faint end of an \Ha\ LF.
The extinction is broadly correlated with SFR, stellar mass, and \Ha\ equivalent width \citep[see also][]{Garn10,GarnBest10,Ly12,Dominguez12,Kashino13}.
To address which parameter is more fundamental in governing extinction is beyond the scope of this work. 

{\it We argue that \Ha\ LF at $z=2.24$ mirrors the SMFs of SFGs at $z\sim 2$}.
We know that the SED-derived extinction tends to be underestimated.
Therefore, our \Ha\ LF is likely underestimated too.
Taking this into account, the agreement between our \Ha\ LF and these converted from SMFs is notably good. 
The evolution of SMFs or cosmic variance may also induce additional uncertainties.
We emphasize that this finding is indeed a reflection of the main sequence of SFGs.  
The excellent agreement between the \Ha\ LF and the SMFs confirms that the main sequence has a slope close to unity. 
As a consequence, the \Ha-selection and the color-selection target on the same population of SFGs and naturally link the two branches of studies together.

On the other hand, the faint-end slope of \Ha\ LF signifies the relative contribution between galaxies with weak and luminous \Ha, a measurement that is important to the understanding of star formation behavior over cosmic time \citep[e.g.,][]{Dutton10}.
The faint-end of an LF is shaped by processes related to star formation, e.g., feedback from supernovae \citep{Dekel86}, radiatively driven winds \citep{Springel03}, and photons ionizing neutral gas \citep{Kravtsov04}. 
The dependence of the faint-end slope of \Ha\ LF on extinction correction hints that UV LFs at high-$z$ might be in a similar situation.
For instance, \citet{Reddy09} presented the UV LF at $z\sim 2-3$ with a faint-end slope of $-1.73$.
Since UV light is closely coupled with \Ha\ emission, we suspect that the faint-end of the UV LF would become shallower if careful extinction correction is taken into account.

\acknowledgments

We are grateful to the referee for his/her valuable comments and suggestions, which have improved our manuscript.
F.X.A acknowledges Cong Ma for his kind assistance.
This research uses data obtained through the Telescope Access Program (TAP), 
which is funded by the National Astronomical Observatories and the Special 
Fund for Astronomy from the Ministry of Finance.
This work is supported by the Strategic Priority Research Program "The Emergence of Cosmological Structures" of the Chinese Academy of Sciences (Grant No. XDB09000000) and National Basic Research Program of China (973 Program 2013CB834900). W-H.W. acknowledges the grant of National Science Council of Taiwan (102-2119-M-001-007-MY3). X.K. is supported by the National Natural Science Foundation of China (NSFC, Nos. 11225315 and 11320101002) and the Specialized Research Fund for the Doctoral Program of Higher Education (SRFDP, No. 20123402110037).
%





\end{document}

%% file: table1.tex
\begin{table*}
\tabcolsep=9pt
\rotatebox{90}{
\caption{\label{table1} Multi-band Photometry of Our Sample of \Ha\ ELGs}
}
\begin{sideways} 
\tiny \begin{tabular}{cccccccccccccc}
\\[0.3mm]
\hline \\[-2.3mm]
\hline \\[-2.1mm]
ID & Mag\_$U$ & Mag\_$B$ & Mag\_$V$ & Mag\_$V_\mathrm{606}$ & Mag\_$R$ & Mag\_$I$ & Mag\_$z_\mathrm{850}$  & Mag\_$J_\mathrm{125}$ & Mag\_J & Mag\_$H_\mathrm{160}$ & Mag\_$K_\mathrm{s}$ & Mag\_$H_2S1$ & Mag\_$24\mu$m\\

\hline \\[-2mm]
   1      &25.62$\pm$0.18    &24.56$\pm$0.02    &24.34$\pm$0.03    &24.28$\pm$0.02    &24.23$\pm$0.02    &24.41$\pm$0.15    &24.16$\pm$0.03    &23.76$\pm$0.02    &23.69$\pm$0.05    &23.17$\pm$0.01    &23.11$\pm$0.04    &21.89$\pm$0.05                                     &-99.99             \\
   2     &-99.99         \tablenotemark{a}    &25.29$\pm$0.06    &24.93$\pm$0.06    &24.72$\pm$0.03    &24.52$\pm$0.04    &24.29$\pm$0.18    &23.94$\pm$0.03    &22.92$\pm$0.01    &22.96$\pm$0.03    &22.24$\pm$0.01    &21.72$\pm$0.02    &21.18$\pm$0.04                    &18.39 $\pm$0.21    \\
   3     &-99.99             &25.32$\pm$0.05    &25.15$\pm$0.06    &25.06$\pm$0.02    &24.98$\pm$0.04   &-99.99             &24.68$\pm$0.03    &24.22$\pm$0.02    &24.09$\pm$0.06    &23.73$\pm$0.02    &23.28$\pm$0.04    &22.43$\pm$0.07                                     &-99.99             \\
   4     &-99.99            &-99.99             &25.90$\pm$0.11    &25.84$\pm$0.05    &25.79$\pm$0.09   &-99.99             &25.40$\pm$0.07    &24.85$\pm$0.04    &24.65$\pm$0.11    &24.16$\pm$0.02    &23.65$\pm$0.07    &22.84$\pm$0.12                                     &-99.99             \\
   5      &24.17$\pm$0.03    &23.45$\pm$0.01    &23.37$\pm$0.01    &23.37$\pm$0.01    &23.37$\pm$0.01    &23.31$\pm$0.06    &23.38$\pm$0.01    &23.26$\pm$0.01    &23.21$\pm$0.03    &22.73$\pm$0.01    &22.77$\pm$0.03    &21.86$\pm$0.05                                     &-99.99             \\
   6            &...\tablenotemark{b}               &...               &...         &25.19$\pm$0.03          &...\tablenotemark{c}               &...         &26.25$\pm$0.17    &24.62$\pm$0.03    &24.51$\pm$0.10    &23.90$\pm$0.02    &24.20$\pm$0.12    &22.80$\pm$0.12   &-99.99             \\
   7      &25.34$\pm$0.08    &24.75$\pm$0.03    &24.69$\pm$0.04    &24.65$\pm$0.02    &24.60$\pm$0.03    &24.19$\pm$0.13    &24.34$\pm$0.03    &23.76$\pm$0.01    &23.79$\pm$0.04    &23.58$\pm$0.01    &23.32$\pm$0.05    &22.63$\pm$0.11                                     &-99.99             \\
   8      &26.34$\pm$0.18    &25.39$\pm$0.05    &25.14$\pm$0.05    &25.09$\pm$0.02    &25.04$\pm$0.05    &24.96$\pm$0.24    &24.55$\pm$0.03    &23.86$\pm$0.01    &23.80$\pm$0.04    &23.45$\pm$0.01    &22.89$\pm$0.03    &22.53$\pm$0.09                                     &-99.99             \\
   9      &25.68$\pm$0.12    &24.23$\pm$0.02    &23.94$\pm$0.02    &23.92$\pm$0.01    &23.90$\pm$0.02    &23.60$\pm$0.10    &23.40$\pm$0.01    &22.72$\pm$0.01    &22.66$\pm$0.02    &22.30$\pm$0.01    &21.79$\pm$0.02    &21.09$\pm$0.04                                     &19.41 $\pm$0.32    \\
  10      &25.55$\pm$0.08    &24.83$\pm$0.03    &24.74$\pm$0.03    &24.78$\pm$0.02    &24.82$\pm$0.04    &24.75$\pm$0.19    &24.71$\pm$0.03    &24.53$\pm$0.02    &24.38$\pm$0.08    &24.10$\pm$0.02    &24.14$\pm$0.11    &22.74$\pm$0.11                                     &-99.99             \\
  11      &25.79$\pm$0.23    &24.78$\pm$0.03    &24.62$\pm$0.03    &24.64$\pm$0.02    &24.65$\pm$0.03    &24.61$\pm$0.19    &24.58$\pm$0.03    &24.37$\pm$0.02    &24.13$\pm$0.06    &23.77$\pm$0.01    &23.65$\pm$0.07    &22.50$\pm$0.09                                     &-99.99             \\
  12      &25.45$\pm$0.09    &24.50$\pm$0.02    &24.31$\pm$0.03    &24.33$\pm$0.01    &24.34$\pm$0.02    &24.33$\pm$0.15    &24.21$\pm$0.02    &23.70$\pm$0.01    &23.56$\pm$0.04    &23.40$\pm$0.01    &23.06$\pm$0.04    &22.41$\pm$0.08                                     &-99.99             \\
  13      &25.21$\pm$0.07    &24.42$\pm$0.02    &24.42$\pm$0.03    &24.45$\pm$0.01    &24.48$\pm$0.03    &24.46$\pm$0.16    &24.53$\pm$0.03    &24.55$\pm$0.02    &24.58$\pm$0.08    &23.88$\pm$0.01    &23.83$\pm$0.08    &22.23$\pm$0.07                                     &-99.99             \\
  14     &-99.99             &25.95$\pm$0.09    &25.81$\pm$0.10    &25.77$\pm$0.04    &25.74$\pm$0.09   &-99.99             &25.67$\pm$0.08    &25.21$\pm$0.04    &25.06$\pm$0.13    &24.37$\pm$0.02    &24.80$\pm$0.18    &22.99$\pm$0.13                                     &-99.99             \\
  15\tablenotemark{*}      &24.60$\pm$0.20    &23.57$\pm$0.01    &23.40$\pm$0.01    &23.36$\pm$0.01    &23.33$\pm$0.01    &23.18$\pm$0.07    &23.06$\pm$0.01    &22.54$\pm$0.00    &22.60$\pm$0.02    &22.17$\pm$0.00    &21.68$\pm$0.02    &20.83$\pm$0.03                    &18.59 $\pm$0.23    \\
  16\tablenotemark{*}      &25.05$\pm$0.08    &23.96$\pm$0.02    &23.64$\pm$0.02    &23.46$\pm$0.01    &23.28$\pm$0.01    &22.88$\pm$0.05    &22.68$\pm$0.01    &21.91$\pm$0.00    &21.88$\pm$0.01    &21.29$\pm$0.00    &20.73$\pm$0.01    &19.81$\pm$0.01                    &17.82 $\pm$0.16    \\
  17      &24.77$\pm$0.06    &24.11$\pm$0.02    &23.85$\pm$0.02          &...         &23.74$\pm$0.02    &23.60$\pm$0.11          &...         &23.37$\pm$0.01    &22.94$\pm$0.03    &22.82$\pm$0.01    &22.41$\pm$0.03    &21.10$\pm$0.03                                     &19.67 $\pm$0.35    \\
  18     &-99.99            &-99.99            &-99.99             &26.51$\pm$0.11   &-99.99            &-99.99             &25.11$\pm$0.06    &23.80$\pm$0.01    &23.84$\pm$0.07    &23.08$\pm$0.01    &22.36$\pm$0.03    &21.95$\pm$0.07                                     &-99.99             \\
  19      &25.86$\pm$0.15    &24.85$\pm$0.04    &24.64$\pm$0.04    &24.50$\pm$0.02    &24.36$\pm$0.03    &24.06$\pm$0.15    &23.85$\pm$0.03    &23.02$\pm$0.01    &22.91$\pm$0.03    &22.49$\pm$0.01    &21.94$\pm$0.02    &21.24$\pm$0.04                                     &19.05 $\pm$0.28    \\
  20     &-99.99            &-99.99            &-99.99             &25.45$\pm$0.04   &-99.99            &-99.99             &24.72$\pm$0.04    &23.83$\pm$0.01    &24.18$\pm$0.07    &23.21$\pm$0.01    &22.73$\pm$0.03    &21.89$\pm$0.05                                     &19.71 $\pm$0.36    \\
  21      &25.83$\pm$0.11    &25.13$\pm$0.04    &25.00$\pm$0.05    &24.95$\pm$0.02    &24.90$\pm$0.04    &24.71$\pm$0.20    &24.92$\pm$0.04    &24.23$\pm$0.01    &24.16$\pm$0.06    &23.77$\pm$0.01    &23.69$\pm$0.06    &22.48$\pm$0.08                                     &-99.99             \\
  22      &25.11$\pm$0.15    &24.12$\pm$0.02    &23.85$\pm$0.02    &23.77$\pm$0.01    &23.69$\pm$0.02    &23.55$\pm$0.09    &23.31$\pm$0.01    &22.76$\pm$0.00    &22.72$\pm$0.03    &22.40$\pm$0.00    &22.13$\pm$0.02    &21.46$\pm$0.05                                     &-99.99             \\
  23      &24.00$\pm$0.01    &23.25$\pm$0.02    &23.22$\pm$0.01    &23.20$\pm$0.01    &23.18$\pm$0.02    &23.27$\pm$0.05    &23.11$\pm$0.01    &22.81$\pm$0.01    &22.70$\pm$0.03    &22.44$\pm$0.01    &22.35$\pm$0.03    &21.98$\pm$0.07                                     &-99.99             \\
  24      &25.10$\pm$0.07    &24.44$\pm$0.02    &24.24$\pm$0.03    &24.15$\pm$0.01    &24.06$\pm$0.03    &23.78$\pm$0.10    &23.46$\pm$0.02    &22.64$\pm$0.01    &22.63$\pm$0.03    &22.14$\pm$0.00    &21.69$\pm$0.02    &21.25$\pm$0.04                                     &-99.99             \\
  25      &25.37$\pm$0.28    &24.41$\pm$0.02    &24.27$\pm$0.02    &24.25$\pm$0.01    &24.22$\pm$0.02    &24.18$\pm$0.13    &24.12$\pm$0.02    &23.74$\pm$0.01    &23.78$\pm$0.08    &23.44$\pm$0.01    &23.41$\pm$0.07    &22.64$\pm$0.12                                     &-99.99             \\
  26      &25.09$\pm$0.06    &24.48$\pm$0.02    &24.47$\pm$0.03    &24.47$\pm$0.01    &24.48$\pm$0.03    &24.20$\pm$0.13    &24.19$\pm$0.02    &23.66$\pm$0.01    &23.69$\pm$0.04    &23.65$\pm$0.01    &23.38$\pm$0.05    &22.91$\pm$0.12                                     &-99.99             \\
  27      &24.57$\pm$0.05    &23.81$\pm$0.02    &23.56$\pm$0.01    &23.57$\pm$0.01    &23.58$\pm$0.02   &-99.99             &23.75$\pm$0.04          &...         &23.37$\pm$0.04          &...         &22.90$\pm$0.04    &22.41$\pm$0.01                                     &-99.99             \\
  28      &26.23$\pm$0.14    &25.55$\pm$0.05    &25.47$\pm$0.07    &25.39$\pm$0.05    &25.32$\pm$0.06    &24.90$\pm$0.22          &...               &...         &24.85$\pm$0.14          &...         &24.33$\pm$0.14    &22.92$\pm$0.13                                     &-99.99             \\
  29      &25.56$\pm$0.10    &24.96$\pm$0.04    &24.86$\pm$0.04    &24.77$\pm$0.03    &24.67$\pm$0.03    &24.38$\pm$0.16          &...               &...         &23.80$\pm$0.06          &...         &23.41$\pm$0.07    &22.71$\pm$0.12                                     &-99.99             \\
  30     &-99.99             &26.68$\pm$0.15    &26.30$\pm$0.14    &26.29$\pm$0.10    &26.27$\pm$0.14   &-99.99             &25.79$\pm$0.21          &...         &24.66$\pm$0.11          &...         &23.55$\pm$0.06    &23.09$\pm$0.14                                     &-99.99             \\
  31      &25.26$\pm$0.07    &24.53$\pm$0.02    &24.45$\pm$0.03    &24.35$\pm$0.02    &24.26$\pm$0.02    &24.03$\pm$0.11    &24.21$\pm$0.05          &...         &23.77$\pm$0.05          &...         &23.48$\pm$0.06    &22.69$\pm$0.10                                     &-99.99             \\
  32      &25.32$\pm$0.37    &24.19$\pm$0.02    &23.90$\pm$0.02    &23.85$\pm$0.02    &23.80$\pm$0.02    &23.64$\pm$0.11    &23.38$\pm$0.04          &...         &22.88$\pm$0.04          &...         &22.05$\pm$0.03    &20.98$\pm$0.04                                     &18.76 $\pm$0.25    \\
  33      &25.15$\pm$0.23    &24.04$\pm$0.03    &23.82$\pm$0.02    &23.91$\pm$0.02    &24.00$\pm$0.05    &23.69$\pm$0.07    &23.85$\pm$0.05          &...         &23.06$\pm$0.04          &...         &22.04$\pm$0.02    &21.62$\pm$0.05                                     &-99.99             \\
  34      &25.34$\pm$0.08    &24.44$\pm$0.02    &24.18$\pm$0.02    &24.19$\pm$0.02    &24.19$\pm$0.02    &24.06$\pm$0.12    &24.01$\pm$0.05          &...         &23.24$\pm$0.03          &...         &22.76$\pm$0.03    &21.88$\pm$0.04                                     &-99.99             \\
  35     &-99.99             &26.45$\pm$0.18    &26.15$\pm$0.18    &25.91$\pm$0.10    &25.68$\pm$0.11    &25.18$\pm$0.39    &25.03$\pm$0.14          &...         &23.42$\pm$0.05          &...         &21.99$\pm$0.02    &21.59$\pm$0.05                                     &19.34 $\pm$0.31    \\
  36     &-99.99             &25.54$\pm$0.06    &25.26$\pm$0.06    &25.13$\pm$0.07    &25.00$\pm$0.04    &24.83$\pm$0.22    &24.31$\pm$0.11          &...         &23.50$\pm$0.04          &...         &22.82$\pm$0.04    &22.23$\pm$0.07                                     &-99.99             \\
  37     &-99.99            &-99.99             &26.51$\pm$0.18    &26.08$\pm$0.09    &25.65$\pm$0.08   &-99.99             &25.34$\pm$0.14          &...         &24.25$\pm$0.08          &...         &23.07$\pm$0.04    &22.50$\pm$0.08                                     &-99.99             \\
  38      &25.59$\pm$0.12    &24.71$\pm$0.04    &24.47$\pm$0.04    &24.39$\pm$0.03    &24.32$\pm$0.03    &24.39$\pm$0.20    &23.98$\pm$0.06          &...         &23.19$\pm$0.04          &...         &22.47$\pm$0.04    &22.03$\pm$0.08                                     &19.48 $\pm$0.33    \\
  39     &-99.99             &25.11$\pm$0.06    &24.84$\pm$0.05    &24.70$\pm$0.03    &24.56$\pm$0.06   &-99.99             &24.21$\pm$0.06          &...         &23.70$\pm$0.06          &...         &23.06$\pm$0.05    &22.49$\pm$0.09                                     &-99.99             \\
  40      &25.44$\pm$0.12    &24.70$\pm$0.04    &24.48$\pm$0.04    &24.43$\pm$0.03    &24.38$\pm$0.03    &24.15$\pm$0.17    &24.15$\pm$0.07          &...         &23.40$\pm$0.05          &...         &22.74$\pm$0.04    &21.87$\pm$0.07                                     &-99.99             \\
  41     &-99.99             &24.92$\pm$0.03    &24.70$\pm$0.04    &24.66$\pm$0.02    &24.62$\pm$0.03    &24.37$\pm$0.15    &24.27$\pm$0.06          &...         &23.72$\pm$0.04          &...         &23.04$\pm$0.04    &22.03$\pm$0.06                                     &-99.99             \\
  42      &26.11$\pm$0.16    &25.26$\pm$0.05    &25.09$\pm$0.05    &25.16$\pm$0.04    &25.22$\pm$0.05    &24.75$\pm$0.22    &24.66$\pm$0.08          &...         &23.70$\pm$0.06          &...         &23.42$\pm$0.06    &22.81$\pm$0.12                                     &-99.99             \\
  43\tablenotemark{*}      &24.70$\pm$0.06    &23.80$\pm$0.02    &23.65$\pm$0.02    &23.59$\pm$0.01    &23.53$\pm$0.02    &23.22$\pm$0.08    &23.10$\pm$0.03          &...         &22.30$\pm$0.02          &...         &21.14$\pm$0.01    &20.67$\pm$0.02                    &17.73 $\pm$0.16    \\
  44     &-99.99            &-99.99            &-99.99            &-99.99            &-99.99            &-99.99             &25.87$\pm$0.21          &...         &25.14$\pm$0.17          &...         &23.43$\pm$0.05    &22.71$\pm$0.10                                     &-99.99             \\
  45     &-99.99             &25.16$\pm$0.05    &25.03$\pm$0.05    &25.05$\pm$0.04    &25.07$\pm$0.05    &24.59$\pm$0.19    &24.50$\pm$0.07          &...         &23.73$\pm$0.06          &...         &23.29$\pm$0.06    &22.69$\pm$0.11                                     &-99.99             \\
  46     &-99.99             &25.26$\pm$0.07    &24.89$\pm$0.05    &24.87$\pm$0.03    &24.84$\pm$0.07    &24.48$\pm$0.11    &24.30$\pm$0.05          &...         &23.95$\pm$0.07          &...         &23.23$\pm$0.05    &22.29$\pm$0.07                                     &-99.99             \\
  47      &25.25$\pm$0.07    &24.92$\pm$0.03    &24.87$\pm$0.04    &24.86$\pm$0.03    &24.85$\pm$0.04    &24.67$\pm$0.19    &24.45$\pm$0.07          &...         &23.87$\pm$0.06          &...         &23.39$\pm$0.06    &22.81$\pm$0.11                                     &-99.99             \\
  48     &-99.99             &25.86$\pm$0.07    &25.77$\pm$0.09    &25.69$\pm$0.06    &25.62$\pm$0.08   &-99.99             &24.82$\pm$0.09          &...         &24.05$\pm$0.08          &...         &22.77$\pm$0.04    &22.07$\pm$0.07                                     &-99.99             \\
  49      &24.99$\pm$0.06    &24.77$\pm$0.03    &24.79$\pm$0.04    &24.60$\pm$0.02    &24.41$\pm$0.03    &24.41$\pm$0.16    &24.27$\pm$0.06          &...         &24.57$\pm$0.13          &...         &24.08$\pm$0.12    &22.82$\pm$0.13                                     &-99.99             \\
  50      &25.57$\pm$0.08    &24.51$\pm$0.02    &24.27$\pm$0.02    &24.23$\pm$0.02    &24.20$\pm$0.02    &24.00$\pm$0.10    &24.08$\pm$0.04          &...         &23.41$\pm$0.04          &...         &22.82$\pm$0.03    &21.85$\pm$0.05                                     &-99.99             \\
  51      &25.04$\pm$0.06    &24.46$\pm$0.09    &24.28$\pm$0.07    &24.01$\pm$0.02    &23.76$\pm$0.07   &-99.99             &23.52$\pm$0.04          &...         &22.82$\pm$0.03          &...         &21.99$\pm$0.02    &21.28$\pm$0.04                                     &-99.99             \\
  52      &25.36$\pm$0.11    &24.39$\pm$0.03    &24.18$\pm$0.03    &24.12$\pm$0.02    &24.07$\pm$0.03    &24.04$\pm$0.15    &23.94$\pm$0.05          &...         &23.17$\pm$0.06          &...         &22.38$\pm$0.04    &21.29$\pm$0.04                                     &20.18 $\pm$0.44    \\
  53     &-99.99             &25.90$\pm$0.07    &25.59$\pm$0.07    &25.56$\pm$0.05    &25.54$\pm$0.07    &25.21$\pm$0.29    &25.21$\pm$0.13          &...         &23.85$\pm$0.06          &...         &23.15$\pm$0.05    &22.45$\pm$0.08                                     &-99.99             \\
  54     &-99.99             &25.58$\pm$0.07    &25.32$\pm$0.06    &25.32$\pm$0.04    &25.31$\pm$0.06   &-99.99             &25.30$\pm$0.14          &...         &24.34$\pm$0.11          &...         &23.93$\pm$0.10    &22.94$\pm$0.14                                     &-99.99             \\
  55      &25.42$\pm$0.09    &24.89$\pm$0.03    &24.88$\pm$0.04    &24.91$\pm$0.03    &24.93$\pm$0.04    &24.77$\pm$0.22    &24.97$\pm$0.10          &...         &24.84$\pm$0.17          &...         &24.43$\pm$0.15    &22.74$\pm$0.13                                     &-99.99             \\
  56      &24.48$\pm$0.05    &23.48$\pm$0.01    &23.23$\pm$0.01    &23.16$\pm$0.01    &23.10$\pm$0.01    &22.92$\pm$0.06    &22.85$\pm$0.01          &...         &22.25$\pm$0.03          &...         &21.61$\pm$0.02    &20.77$\pm$0.03                                     &-99.99             \\
\hline

\end{tabular}
\\[1.5mm]
\end{sideways}
\rotatebox{90}{
\tiny
$^{a}$Sources are not detected.  $^{b}$Sources could not be resolved in MUSYC observation.  $^{c}$Sources are not covered by CANDELS or GEMS observation. $^{*}$Three X-ray sources.
}

\end{table*}

%% file: table2.tex
\begin{table}[!h]
\begin{center}
\caption{\label{table2} Physical Parameters of Our Sample of \Ha\ ELGs}
\tiny 
\begin{tabular}{ccccccc}
\\[0.3mm]
\hline \\[-2.3mm]
\hline \\[-2.1mm]
ID & Photo-$z$ & Spec-$z$ & A(H$\alpha$)~(mag) & log~H$\alpha$~(erg~s$^{-1}$) & log~$(M/\Msun)$ & Morphology\tablenotemark{a} \\
\hline \\[-2mm]
    1      &2.25     & ...     &1.07    &42.99     & 9.95&  1 \\
    2      &2.24     & ...     &1.92    &43.39     &10.75 & 2  \\ 
    3      &2.24     & ...     &1.70    &42.94     & 9.42&  4  \\ 
    4      &2.24     & ...     &2.00    &42.87     & 9.54&  5  \\ 
    5      &2.25     &2.21     &0.65    &42.76     & 9.33&  2  \\ 
    6      &2.24     &2.22     &0.00    &42.22     & 9.81&  1  \\  
    7      &2.24     & ...     &0.36    &42.25     & 9.81&  1  \\ 
    8      &2.24     & ...     &1.40    &42.52     &10.15 & 2  \\ 
    9      &2.25     &2.23     &1.06    &43.17     &10.62 & 3  \\ 
   10      &2.25     & ...     &0.67    &42.52     & 9.21&  4  \\ 
   11      &2.25     & ...     &1.41    &42.86     & 9.61&  2  \\
   12      &2.25     & ...     &0.64    &42.46     & 9.91&  2  \\
   13      &2.25     &2.22     &0.00    &42.46     & 9.35&  2  \\ 
   14      &2.24     & ...     &0.00    &42.19     & 9.14&  3  \\ 
   15\tablenotemark{*}      &2.24     &2.23     &1.67    &43.55     &10.69 & 2  \\ 
   16\tablenotemark{*}      &2.24     &2.22     &1.73    &44.00     &11.36 & 2  \\ 
   17      &2.24     & ...     &1.58    &43.47     &10.35 & 3  \\ 
   18      &2.24     & ...     &2.10    &42.99     &10.53 & 5  \\ 
   19      &2.24     &2.34     &1.70    &43.30     &10.74 & 3  \\ 
   20      &2.24     & ...     &2.10    &43.31     &10.57 & 3  \\ 
   21      &2.24     &2.28     &0.95    &42.70     & 9.76 & 4  \\
   22      &2.24     &2.23     &0.53    &42.70     &10.23 & 3  \\ 
   23      &2.26     &2.26     &0.03    &42.28     &10.14 & 2  \\ 
   24      &2.24     &2.25     &1.62    &43.14     &10.93 & 3  \\ 
   25      &2.25     &2.15     &0.04    &42.17     & 9.50 & 4  \\ 
   26      &2.23     & ...     &0.04    &41.91     & 9.36 & 4  \\ 
   27      &2.26     & ...     &0.16    &42.17     & 9.98 & 1  \\ 
   28      &2.25     & ...     &0.93    &42.55     & 9.45 & 4  \\
   29      &2.24     & ...     &0.05    &42.12     & 9.87 & 3  \\ 
   30      &2.24     & ...     &1.83    &42.53     & 9.82 & 5  \\ 
   31      &2.25     & ...     &0.01    &42.13     & 9.39 & 1  \\ 
   32      &2.25     &2.21     &1.32    &43.43     &10.49 & 2  \\ 
   33      &2.24     &1.83     &1.81    &43.11     &10.53 & 1  \\ 
   34      &2.24     & ...     &0.36    &42.63     &10.21 & 1  \\ 
   35      &2.24     & ...     &1.93    &43.14     &10.91 & 5  \\ 
   36      &1.82     & ...     &1.33    &42.77     & 9.94 & 4  \\ 
   37      &2.24     & ...     &1.87    &42.86     &10.17 & 5  \\ 
   38      &2.24     & ...     &1.05    &42.58     &10.38 & 3  \\ 
   39      &2.25     & ...     &0.18    &42.19     & 9.89 & 5  \\ 
   40      &2.24     & ...     &1.14    &42.75     &10.22 & 2  \\ 
   41      &2.24     & ...     &1.40    &43.01     & 9.86 & 4  \\ 
   42      &1.83     & ...     &0.18    &42.08     & 9.51 & 2  \\ 
   43\tablenotemark{*}       &2.25     & ...     &1.77    &43.50     &11.14 & 4  \\ 
   44      &2.25     & ...     &1.06    &42.52     &10.23 & 5  \\ 
   45      &1.82     & ...     &0.00    &42.06     & 9.86 & 3  \\ 
   46      &2.24     &1.87     &1.08    &42.77     & 9.77 & 3  \\ 
   47      &2.24     & ...     &0.09    &42.03     & 9.83 & 3  \\ 
   48      &2.24     & ...     &2.09    &43.18     &10.58 & 3  \\
   49      &2.24     & ...     &0.54    &42.41     & 8.40 & 1  \\ 
   50      &2.25     &2.36     &0.12    &42.57     &10.02 & 1  \\ 
   51      &2.24     & ...     &1.44    &43.10     &10.71 & 1  \\ 
   52      &2.25     &2.22     &1.32    &43.28     &10.52 & 2  \\ 
   53      &1.83     &2.10     &0.14    &42.24     & 9.91 & 5  \\ 
   54      &2.24     & ...     &1.05    &42.50     & 9.46 & 4  \\ 
   55      &2.24     & ...     &0.42    &42.45     & 9.26 & 1  \\ 
   56      &2.25     &2.23     &0.23    &43.02     &10.64 & 2  \\ 
\hline

\end{tabular}
\\[1.5mm]
{$^{a}$1:Two components; 2:merger; 3: spiral/diffuse/clumpy; 4:compact; 5:UV-faint. $^{*}$Three X-ray sources (corresponded ID = 435, 29 and 399 in the \textit{Chandra} catalog).}
\end{center}

\end{table}

%% file: ms.bbl
\begin{thebibliography}

\bibitem[An et al.(2013)]{An13} An, F.~X., Zheng, X.~Z., Meng,Y.~Z., et al.\ 2013, Sci China-Phys Mech Astron, 56, 2226
\bibitem[Atek et al.(2010)]{Atek10} Atek, H., Malkan, M., McCarthy, P., et al.\ 2010, \apj, 723, 104 
\bibitem[Bertin \& Arnouts(1996)]{Bertin96} Bertin, E., \& Arnouts, S.\ 1996, \aaps, 117, 393 
\bibitem[Bothwell et al.(2011)]{Bothwell11} Bothwell, M.~S., Kenicutt, R.~C., Johnson, B.~D., et al.\ 2011, \mnras, 415, 1815 
\bibitem[Brammer et al.(2008)]{Brammer08} Brammer, G.~B., van Dokkum, P.~G., \& Coppi, P.\ 2008, \apj, 686, 1503 
\bibitem[Brammer et al.(2011)]{Brammer11} Brammer, G.~B., Whitaker, K.~E., van Dokkum, P.~G., et al.\ 2011, \apj, 739, 24 
\bibitem[Brinchmann et al.(2004)]{Brinchmann04} Brinchmann, J., Charlot, S., White, S.~D.~M., et al.\ 2004, \mnras, 351, 1151 
\bibitem[Bunker et al.(1995)]{Bunker95} Bunker, A.~J., Warren, S.~J., Hewett, P.~C., \& Clements, D.~L.\ 1995, \mnras, 273, 513 
\bibitem[Caldwell et al.(2008)]{Caldwell08} Caldwell, J.~A.~R., McIntosh, D.~H., Rix, H.-W., et al.\ 2008, \apjs, 174, 136 
\bibitem[Calzetti et al.(2000)]{Calzetti00} Calzetti, D., Armus, L., Bohlin, R.~C., et al.\ 2000, \apj, 533, 682 
\bibitem[Cardamone et al.(2010)]{Cardamone10} Cardamone, C.~N., van Dokkum, P.~G., Urry, C.~M., et al.\ 2010, \apjs, 189, 270 
\bibitem[Chapman et al.(2005)]{Chapman05} Chapman, S.~C., Blain, A.~W., Smail, I., \& Ivison, R.~J.\ 2005, \apj, 622, 772 
\bibitem[Colbert et al.(2013)]{Colbert13} Colbert, J.~W., Teplitz, H., Atek, H., et al.\ 2013, \apj, 779, 34 
\bibitem[Conselice et al.(2003)]{Conselice03} Conselice, C.~J., Bershady, M.~A., Dickinson, M., \& Papovich, C.\ 2003, \aj, 126, 1183 
\bibitem[Conselice et al.(2011)]{Conselice11} Conselice, C.~J., Bluck, A.~F.~L., Ravindranath, S., et al.\ 2011, \mnras, 417, 2770 
\bibitem[Daddi et al.(2004)]{Daddi04} Daddi, E., Cimatti, A., Renzini, A., et al.\ 2004, \apj, 617, 746 
\bibitem[Daddi et al.(2007)]{Daddi07} Daddi, E., Dickinson, M., Morrison, G., et al.\ 2007, \apj, 670, 156 
\bibitem[Dekel \& Silk(1986)]{Dekel86} Dekel, A., \& Silk, J.\ 1986, \apj, 303, 39 
\bibitem[Dickinson \& FIDEL Team(2007)]{Dickinson07} Dickinson, M., \& FIDEL Team 2007, BAAS, 39, 822 
\bibitem[Dom{\'{\i}}nguez et al.(2013)]{Dominguez12} Dom{\'{\i}}nguez, A., Siana, B., Henry, A.~L., et al.\ 2013, \apj, 763, 145
\bibitem[Dutton et al.(2010)]{Dutton10} Dutton, A.~A., van den Bosch, F.~C., \& Dekel, A.\ 2010, \mnras, 405, 1690 
\bibitem[Elbaz et al.(2007)]{Elbaz07} Elbaz, D., Daddi, E., Le Borgne, D., et al.\ 2007, \aap, 468, 33 
\bibitem[Erb et al.(2006b)]{Erb06b} Erb, D.~K., Steidel, C.~C., Shapley, A.~E., et al.\ 2006, \apj, 646, 107 
\bibitem[Fang et al.(2012)]{Fang12} Fang, G., Kong, X., Chen, Y., \& Lin, X.\ 2012, \apj, 751, 109
\bibitem[Fioc \& Rocca-Volmerange(1997)]{Fioc97} Fioc, M., \& Rocca-Volmerange, B.\ 1997, \aap, 326, 950 
\bibitem[Franx et al.(2003)]{Franx03} Franx, M., Labb{\'e}, I., Rudnick, G., et al.\ 2003, \apj L, 587, L79 
\bibitem[Garn \& Best(2010)]{GarnBest10} Garn, T., \& Best, P.~N.\ 2010, \mnras, 409, 421 
\bibitem[Garn et al.(2010)]{Garn10} Garn, T., Sobral, D., Best, P.~N., et al.\ 2010, \mnras, 402, 2017 
\bibitem[Gawiser et al.(2006)]{Gawiser06} Gawiser, E., van Dokkum, P.~G., Herrera, D., et al.\ 2006, \apjs, 162, 1 
\bibitem[Geach et al.(2008)]{Geach08} Geach, J.~E., Smail, I., Best, P.~N., et al.\ 2008, \mnras, 388, 1473 
\bibitem[Geach et al.(2012)]{Geach12} Geach, J.~E., Sobral, D., Hickox, R.~C., et al.\ 2012, \mnras, 426, 679 
\bibitem[Grogin et al.(2011)]{Grogin11} Grogin, N.~A., Kocevski, D.~D., Faber, S.~M., et al.\ 2011, \apjs, 197, 35 
\bibitem[Guo et al.(2013)]{Guo13} Guo, K., Zheng, X.~Z., \& Fu, H.\ 2013, \apj, 778, 23 
\bibitem[Hatch et al.(2011)]{Hatch11} Hatch, N.~A., Kurk, J.~D., Pentericci, L., et al.\ 2011, \mnras, 415, 2993 
\bibitem[Hayes et al.(2010)]{Hayes10} Hayes, M., Schaerer, D., {\"O}stlin, G.\ 2010, \aap, 509, L5 
\bibitem[Hopkins \& Beacom(2006)]{Hopkins06} Hopkins, A.~M., \& Beacom, J.~F.\ 2006, \apj, 651, 142 
\bibitem[Hsieh et al.(2012)]{Hsieh12} Hsieh, B.-C., Wang, W.-H., Hsieh, C.-C., et al.\ 2012, \apjs, 203, 23 
\bibitem[Ilbert et al.(2010)]{Ilbert10} Ilbert, O., Salvato, M., Le Floc'h, E., et al.\ 2010, \apj, 709, 644 
\bibitem[Karim et al.(2011)]{Karim11} Karim, A., Schinnerer, E., Mart{\'{\i}}nez-Sansigre, A., et al.\ 2011, \apj, 730, 61 
\bibitem[Kashino et al.(2013)]{Kashino13} Kashino, D., Silverman, J.~D., Rodighiero, G., et al.\ 2013, \apj L, 777, L8
\bibitem[Kennicutt \& Evans(2012)]{Kennicutt12} Kennicutt, R.~C., \& Evans, N.~J.\ 2012, \araa, 50, 531 
\bibitem[Kennicutt(1998)]{Kennicutt98} Kennicutt, R.~C., Jr.\ 1998, \araa, 36, 189
\bibitem[Koekemoer et al.(2011)]{Koekemoer11} Koekemoer, A.~M., Faber, S.~M., Ferguson, H.~C., et al.\ 2011, \apjs, 197, 36 
\bibitem[Kravtsov et al.(2004)]{Kravtsov04} Kravtsov, A.~V., Berlind, A.~A., Wechsler, R.~H., et al.\ 2004, \apj, 609, 35 
\bibitem[Kriek et al.(2009)]{Kriek09} Kriek, M., van Dokkum, P.~G., Labb{\'e}, I., et al.\ 2009, \apj, 700, 221 
\bibitem[Kroupa(2001)]{Kroupa01} Kroupa, P.\ 2001, \mnras, 322, 231 
\bibitem[Law et al.(2012)]{Law12} Law, D.~R., Steidel, C.~C., Shapley, A.~E., et al.\ 2012, \apj, 745, 85 
\bibitem[Lee et al.(2007)]{Lee07} Lee, J.~C., Kennicutt, R.~C., Funes, S.~J., et al.\ 2007, \apjl, 671, L113 
\bibitem[Lee et al.(2012)]{Lee12} Lee, J.~C., Ly, C., Spitler, L., et al.\ 2012, \pasp, 124, 782 
\bibitem[Leitherer et al.(1999)]{Leitherer99} Leitherer, C., Schaerer, D., Goldader, J.~D., et al.\ 1999, \apjs, 123, 3 
\bibitem[Lotz et al.(2006)]{Lotz06} Lotz, J.~M., Madau, P., Giavalisco, M., Primack, J., \& Ferguson, H.~C.\ 2006, \apj, 636, 592 
\bibitem[Lotz et al.(2004)]{Lotz04} Lotz, J.~M., Primack, J., \& Madau, P.\ 2004, \aj, 128, 163
\bibitem[Ly et al.(2011)]{Ly11} Ly, C., Lee, J.~C., Dale, D.~A., et al.\ 2011, \apj, 726, 109 
\bibitem[Ly et al.(2012)]{Ly12} Ly, C., Malkan, M.~A., Kashikawa, N., et al.\ 2012, \apj L, 747, L16 
\bibitem[Maraston(2005)]{Maraston05} Maraston, C.\ 2005, \mnras, 362, 799 
\bibitem[Matsuda et al.(2011)]{Matsuda11} Matsuda, Y., Smail, I., Geach, J.~E., et al.\ 2011, \mnras, 416, 2041 
\bibitem[Moorwood et al.(2000)]{Moorwood00} Moorwood, A.~F.~M., van der Werf, P.~P., Cuby, J.~G., \& Oliva, E.\ 2000, \aap, 362, 9 
\bibitem[Noeske et al.(2007)]{Noeske07} Noeske, K.~G., Weiner, B.~J., Faber, S.~M., et al.\ 2007, \apj L, 660, L43 
\bibitem[Oke(1974)]{Oke74} Oke, J.~B.\ 1974, \apjs, 27, 21 
\bibitem[Papovich et al.(2005)]{Papovich05} Papovich, C., Dickinson, M., Giavalisco, M., Conselice, C.~J., \& Ferguson, H.~C.\ 2005, \apj, 631, 101 
\bibitem[Peng et al.(2010)]{Peng10} Peng, Y.~J., Lilly, S.~J., Kova{\v c}, K., et al.\ 2010, \apj, 721, 193 
\bibitem[Puget et al.(2004)]{Puget04} Puget, P., Stadler, E., Doyon, R., et al.\ 2004, \procspie, 5492, 978 
\bibitem[Ravindranath et al.(2006)]{Ravindranath06} Ravindranath, S., Giavalisco, M., Ferguson, H.~C., et al.\ 2006, \apj, 652, 963 
\bibitem[Reddy et al.(2012)]{Reddy12} Reddy, N.~A., Pettini, M., Steidel, C.~C., et al.\ 2012, \apj, 754, 25 
\bibitem[Reddy \& Steidel(2009)]{Reddy09} Reddy, N.~A., \& Steidel, C.~C.\ 2009, \apj, 692, 778 
\bibitem[Renzini(2006)]{Renzini06} Renzini, A.\ 2006, \araa, 44, 141 
\bibitem[Retzlaff et al.(2010)]{Retzlaff10} Retzlaff, J., Rosati, P., Dickinson, M., et al.\ 2010, \aap, 511, A50 
\bibitem[Rix et al.(2004)]{Rix04} Rix, H.-W., Barden, M., Beckwith, S.~V.~W., et al.\ 2004, \apjs, 152, 163 
\bibitem[Schechter(1976)]{Schechter76} Schechter, P.\ 1976, \apj, 203, 297 
\bibitem[Sedgwick et al.(2013)]{Sedgwick13} Sedgwick, C., Serjeant, S., Pearson, C., et al.\ 2013, \mnras, 436, 395
\bibitem[Shapley(2011)]{Shapley11} Shapley, A.~E.\ 2011, \araa, 49, 525 
\bibitem[Sobral et al.(2012)]{Sobral12} Sobral, D., Best, P.~N., Matsuda, Y., et al.\ 2012, \mnras, 420, 1926 
\bibitem[Sobral et al.(2013)]{Sobral13} Sobral, D., Smail, I., Best, P.~N., et al.\ 2013, \mnras, 428, 1128 
\bibitem[Springel \& Hernquist(2003)]{Springel03} Springel, V., \& Hernquist, L.\ 2003, \mnras, 339, 312 
\bibitem[Steidel et al.(1999)]{Steidel99} Steidel, C.~C., Adelberger, K.~L., Giavalisco, M., Dickinson, M., \& Pettini, M.\ 1999, \apj, 519, 1 
\bibitem[Stott et al.(2013)]{Stott13} Stott, J.~P., Sobral, D., Smail, I., et al.\ 2013, \mnras, 430, 1158 
\bibitem[Swinbank et al.(2004)]{Swinbank04} Swinbank, A.~M., Smail, I., Chapman, S.~C., et al.\ 2004, \apj, 617, 64 
\bibitem[van Dokkum et al.(2011)]{vanDokkum11} van Dokkum, P.~G., Brammer, G., Fumagalli, M., et al.\ 2011, \apj L, 743, L15 
\bibitem[Wang et al.(2010)]{Wang10} Wang, W.-H., Cowie, L.~L., Barger, A.~J., Keenan, R.~C., \& Ting, H.-C.\ 2010, \apjs, 187, 251 
\bibitem[Wuyts et al.(2011a)]{Wuyts11a} Wuyts, S., F{\"o}rster Schreiber, N.~M., Lutz, D., et al.\ 2011, \apj, 738, 106 
\bibitem[Wuyts et al.(2011b)]{Wuyts11b} Wuyts, S., F{\"o}rster Schreiber, N.~M., van der Wel, A., et al.\ 2011, \apj, 742, 96 
\bibitem[Xue et al.(2011)]{Xue11} Xue, Y.~Q., Luo, B., Brandt, W.~N., et al.\ 2011, \apjs, 195, 10 
\bibitem[Yan et al.(1999)]{Yan99} Yan, L., McCarthy, P.~J., Freudling, W., et al.\ 1999, \apjl, 519, L47 
\bibitem[Zheng et al.(2006)]{zheng06} Zheng, X.~Z., Bell, E.~F., Rix, H.-W., et al.\ 2006, \apj, 640, 784
\bibitem[Zheng et al.(2007)]{Zheng07} Zheng, X.~Z., Dole, H., Bell, E.~F., et al.\ 2007, \apj, 670, 301
\end{thebibliography}
